\documentclass[lettersize,journal]{IEEEtran}
\IEEEoverridecommandlockouts
\usepackage{cite}
\usepackage{amsmath,amssymb,amsfonts}
\usepackage{amssymb}
\usepackage{algpseudocode}
\usepackage{amsfonts}
\usepackage{graphicx}
\usepackage{fancyhdr}  %
\usepackage{cases}
\usepackage{textcomp}
\usepackage{extarrows}
\usepackage{algorithm,algpseudocode}
\usepackage{multirow,tabularx}
\usepackage{mathtools}
\usepackage{xcolor}
\usepackage[english]{babel}
\usepackage{amsthm}

\def\BibTeX{{\rm B\kern-.05em{\sc i\kern-.025em b}\kern-.08em
		T\kern-.1667em\lower.7ex\hbox{E}\kern-.125emX}}

\usepackage{fancyhdr}

\begin{document}

	\title{Joint Channel Estimation and Signal Recovery for RIS-Empowered Multi-User Communications}
 
	\author{Li Wei, Chongwen Huang, Qinghua Guo,~\IEEEmembership{Senior Member,~IEEE}, Zhaohui Yang,~\IEEEmembership{Member,~IEEE}, Zhaoyang Zhang,~\IEEEmembership{Senior Member,~IEEE},  George~C.~Alexandropoulos,~\IEEEmembership{Senior Member,~IEEE}, M\'{e}rouane~Debbah,~\IEEEmembership{Fellow,~IEEE} and Chau~Yuen,~\IEEEmembership{Fellow,~IEEE}
		\thanks{L. Wei and C. Yuen are with the Engineering Product Development (EPD) Pillar, Singapore University of Technology and Design, Singapore 487372 (e-mails: wei\_li@mymail.sutd.edu.sg,  yuenchau@sutd.edu.sg).}
		
		\thanks{C.~Huang, Z. Yang and Z.~Zhang are with the College of Information Science and Electronic Engineering, Zhejiang University, Hangzhou 310007, China, and Zhejiang Provincial Key Lab of Information Processing, Communication and Networking (IPCAN), Hangzhou 310007, China, and the International Joint Innovation Center, Zhejiang University, Haining 314400, China. C.~Huang is also with Zhejiang-Singapore Innovation and AI Joint Research Lab, Hangzhou 310027, China. Z. Yang is also with the Zhejiang Laboratory, Hangzhou 31121, China.  (e-mails: \{chongwenhuang@zju.edu.cn, zhaohuiyang92@gmail.com,ning\_ming@zju.edu.cn)\}.  }

		\thanks{Q. Guo is with the School of Electrical, Computer and Telecommunications Engineering, University of Wollongong, Wollongong, NSW 2522, Australia (e-mail: qguo@uow.edu.au).}

		\thanks{G.~C.~Alexandropoulos is with the Department of Informatics and Telecommunications,
			National and Kapodistrian University of Athens, Panepistimiopolis Ilissia, 15784 Athens, Greece and also with the Technology Innovation Institute, Abu Dhabi, United Arab Emirates (e-mail: alexandg@di.uoa.gr).}
		 
		\thanks{ M. Debbah is with the Technology Innovation Institute, 9639 Masdar City, Abu Dhabi, United Arab Emirates (email: merouane.debbah@tii.ae) and also with CentraleSupelec, University Paris-Saclay, 91192 Gif-sur-Yvette, France.}
	}

	\maketitle
	\thispagestyle{fancy}
	\begin{abstract}
		Reconfigurable intelligent surfaces (RISs) have been recently considered as a promising candidate for energy-efficient solutions in future wireless networks. Their dynamic and low-power configuration enables coverage extension, massive connectivity, and low-latency communications. Due to a large number of unknown variables referring to the RIS unit elements and the transmitted signals, channel estimation and signal recovery in RIS-based systems are the ones of the most critical technical challenges. To address this problem, we focus on the RIS-assisted wireless communication system and present two joint channel estimation and signal recovery schemes based on message passing algorithms in this paper. Specifically, the proposed bidirectional scheme applies the Taylor series expansion and Gaussian approximation to simplify the sum-product procedure in the formulated problem. In addition, the inner iteration that adopts two variants of approximate message passing algorithms is incorporated to ensure robustness and convergence. Two ambiguities removal methods are also discussed in this paper. Our simulation results show that the proposed schemes show the superiority over the state-of-art benchmark method. We also provide insights on the impact of different RIS parameter settings on the proposed schemes.
	\end{abstract}
	
	
	\section{Introduction}\label{sec:intro}
	Reconfigurable intelligent surface (RIS) is a potential candidate technology for beyond fifth-generation (5G) wireless communications \cite{Akyildiz2018mag,hu2018beyond,huang2019reconfigurable,Marco2019,qingqing2019towards,strinati2021wireless}. A large number of hardware-efficient passive reflecting elements are employed in a RIS to facilitate low-power, energy-efficient, high-speed, massive-connectivity, and low-latency communications  \cite{9140329,9371416,9206044,9136592}. Each unit element in a RIS can alter the phase of the incoming signal without requiring a dedicated power amplifier that is needed in conventional amplify-and-forward relaying systems \cite{alexandropoulos2021reconfigurable,9136592,huang2019reconfigurable}. As a result, RISs had gained much attention in recent years. 
	
	The energy efficiency potential of RIS in the scenario of outdoor multi-user multiple input single output communications was analyzed in \cite{huang2019reconfigurable}, while \cite{husha_LIS2} focused on an indoor scenario to illustrate the potential of RIS-based indoor positioning. It was shown in \cite{9110869} and \cite{han2019} that the potential of positioning is large even in the case of RISs with finite resolution unit elements and statistical channel knowledge. Recently, a novel passive beamforming and information transfer technique was proposed in \cite{8941126} to enhance primary communications, as well as a two-step approach at the receiver to retrieve the information from both the transmitter and RIS. RIS-assisted communications in the millimeter-wave  and terahertz bands were also lately investigated to deal with limited transmission distance problem  \cite{Akyildiz2018mag}. Orthogonal and non-orthogonal multiple access in RIS-assisted communications were studied in \cite{9133094} as   cost-effective solutions for boosting spectrum/energy efficiency.  {The work in \cite{9318531} proposed an RIS aided multi-user full-duplex two-way communication network to enhance user fairness, where RIS is effectively adjusted to mitigate the interference at the users with fully knowledge of channel information. The work in  \cite{9263383} focused on the RIS aided two-way device-to-device multi-pair OFDM communications and maximized the minimum bidirectional weighted sum-rate with perfect CSI.} The existing research works have proved the great potential of RISs, however, most of the existing research works focusing on RIS configuration optimization were carried out with the assumption of  the perfect channel state information. Thus, channel acquisition plays a significant role in RIS-assisted systems, and it is a challenging issue due to the incorporation of a large number of passive elements in RISs, specifically, the estimation of the cascaded channels, transmitter-to-RIS channel and RIS-to-receiver channel, brings many difficulties.  

	The recent works in \cite{Alkhateeb2019} and \cite{9410457} presented compressive sensing and deep learning approaches for recovering the involved channels and designing the RIS phase matrix. However, the deep learning approaches require extensive training during offline training phases, and the compressive sensing framework assumes that RIS has some active elements, i.e., a fully digital or hybrid analog and digital architecture is attached at RIS. The architecture that has some active elements would increase the RIS hardware complexity and power consumption. A low-power radio frequency chain for channel estimation was considered in \cite{qingqing2019towards}, which also requires additional energy consumption compared to passive RISs. Recently, the authors in  \cite{9366805} adopted PARAllel FACtor (PARAFAC) decomposition to estimate all involved channels with totally passive RIS elements, but the PARAFAC based methods require strict feasibility conditions with the excessively high complexity for a large number of RIS elements. In \cite{9133142}, the authors adopted a hierarchical training reflection design to progressively estimate channels over multiple time blocks, which has high complexity in the multiuser communication system. However, most of the existing works only consider the channel estimation part in RIS-assisted systems, while the joint channel estimation and signal recovery problem has not been well studied. 
	
     In \cite{9417121}, the authors adopted a two-stage approach to estimate channels and transmitted signal separately. Specifically, the cascaded channels are estimated using the PARAFAC method firstly, and then the signal is recovered based on the estimated channels. The existing literature applies similar methods, i.e., the channel information is obtained in the first stage, and then the signal is recovered in the second stage. Such a method cannot fully explore the characteristics of channels and transmitted signals due to separate channel estimation and signal recovery, and the training overhead is high in this two-stage method. Thus, some  message passing algorithms based on factor graphs were proposed for joint estimation and signal recovery, such as bilinear generalized approximate message passing (BiGAMP) \cite{6898015} and generalized sparse Bayesian learning algorithm \cite{8357527}.  In \cite{7574287}, the framework of expectation propagation was employed to perform joint channel estimation and decoding in massive multiple-input multiple-output (MIMO) systems with orthogonal frequency division multiplexing. In \cite{9103622}, the authors designed a Bayesian method for the effective channel estimation and signal recovery in grant-free non-orthogonal multiple access. In \cite{9006927}, the authors applied a new expectation maximization message passing algorithm combination for joint channel estimation and symbol detection. In \cite{zou2020multi}, the authors proposed a multi-layer BiGAMP approach for handling the cascaded problem in amplify-and-forward relay communication systems, and the performance of joint channel estimation and signal detection in two-hop relay systems with one known channel was investigated. However, these approximate message passing (AMP) related algorithms are vulnerable to ill-conditioned measurement matrices that may cause divergence. Thus, some variants to improve this problem were proposed, such as damping method \cite{6998861},  bilinear adaptive generalized vector approximate message passing \cite{8580585} and  AMP with unitary transformation (UTAMP) \cite{9293406,guo2015approximate}. The above works in \cite{6998861,8580585,9293406,guo2015approximate} provide some insights into the joint channel estimation and signal recovery in RIS-assisted communication systems, which motivates us to explore a new technique to reliably estimate channels and recover signal simultaneously with tolerable training overhead.
	
	In this paper, we propose a novel joint channel estimation and signal detection algorithm in a RIS-assisted  {uplink} wireless communication system, where a multi-antenna base station (BS) serves multiple single-antenna users.  {Different from the existing methods, which estimate channels (with fixed number of pilots) and detect signal in two separate stages, we take advantage of iterative process to estimate channels and transmitted signal simultaneously, specifically, the estimated data in each iteration is used as virtual training signals to achieve better performance.} Specifically, we formulate the joint channel estimation and signal recovery as an inference problem that estimates two cascaded channels and transmitted signal simultaneously. The factor graph and the related sum-product message passing rules of the formulated problem are developed, then we apply Taylor series expansion and Gaussian approximation to deal with the tricky inference problem.  We also propose two schemes that improve the convergency by iteratively refining the estimates of the proposed algorithm.  Moreover, to solve the issue of involved ambiguities in the formulated problem, two ambiguity removal approaches are investigated in this paper.  The proposed algorithms  are efficient and provide good channel estimation and signal recovery performance. Our extensive simulation results validate the effectiveness of the proposed techniques and their favorable performance. The main contributions of this paper are summarized as follows:
	\begin{enumerate}
		\item  A proper system model of the joint channel estimation and signal detection in a RIS-assisted wireless system is established, and the factor  representation is derived as well as the related message passing algorithms. Moreover,  simplification of the message updating for efficient inference is studied in this work. Specifically, inspired by AMP derivations, we adopt Gaussian approximation and Taylor series expansion to derive a computationally efficient algorithm. 
		\item Based on the simplified algorithms, two efficient schemes that further improve the convergence are presented. One is bidirectional AMP (BAMP) and the other is bidirectional unitary transformation AMP (BUTAMP), for robustness and faster convergence. Specifically, all channels and transmitted signals can be recovered without divergence using the proposed two schemes. 
		\item We investigate the inherent ambiguities in the proposed algorithm, and two ambiguities removal methods are proposed. Specifically, the introduction of a large number of pilots could eliminate phase ambiguities and scaling ambiguities (the known parts in estimated channel and transmitted signal are termed as pilots), however, it is impractical to obtain these pilots directly, thus we investigate the efficient methods to obtain these pilots.  In addition, the computational complexity is also analyzed. The proposed algorithms have the low computational complexity,  which is dominated by matrix-vector multiplications. 
		\item We prove the effectiveness of the proposed algorithms through simulations and comparisons with some benchmark curves. It is shown that the proposed schemes can achieve  favorable performance in various cases.
	\end{enumerate}
	
	The remainder of this paper is organized as follows. In Section \ref{sec:format}, the system model is introduced. In Section~\ref{sec:channel_est}, the factor graph and the proposed message passing algorithm  are presented, we propose two schemes to further improve the convergence of the algorithms, and ambiguities removal and complexity analysis are also discussed.   Section \ref{sec:simulation} presents the numerical results of the proposed schemes. Finally, some conclusions are drawn in Section~\ref{sec:conclusion}.
	
	\textit{Notation}: Fonts $a$, $\mathbf{a}$, and $\mathbf{A}$ represent scalars, vectors, and matrices, respectively. We use $\mathbf{A}^T$, $\mathbf{A}^H$, $\mathbf{A}^{-1}$ and $\mathbf{A^\dag}$  to denote the transpose, Hermitian (conjugate transpose), inverse and pseudo-inverse of $ \mathbf{A} $, respectively. The $(m,n)$-th entry of $\mathbf{A}$ is denoted by $a_{mn}$.  We use $|\cdot|$ and $(\cdot)^*$ to denote the modulus and conjugation, respectively.  {$\propto$  denotes the proportional relationship between two quantities.} Finally, notation $diag(\mathbf{a})$ represents a diagonal matrix with the entries of $\mathbf{a}$ on its main diagonal.
	
	\section{System Model}\label{sec:format}
	In this section, we first describe the system and signal models for the considered RIS-assisted wireless communication system, and then derive a message passing algorithm for the end-to-end wireless communication channel.
 
\begin{figure}[!h]\vspace{-0mm}
		\setlength{\belowcaptionskip}{-1cm}  
		\begin{center}
			\centerline{\includegraphics[width=0.5\textwidth]{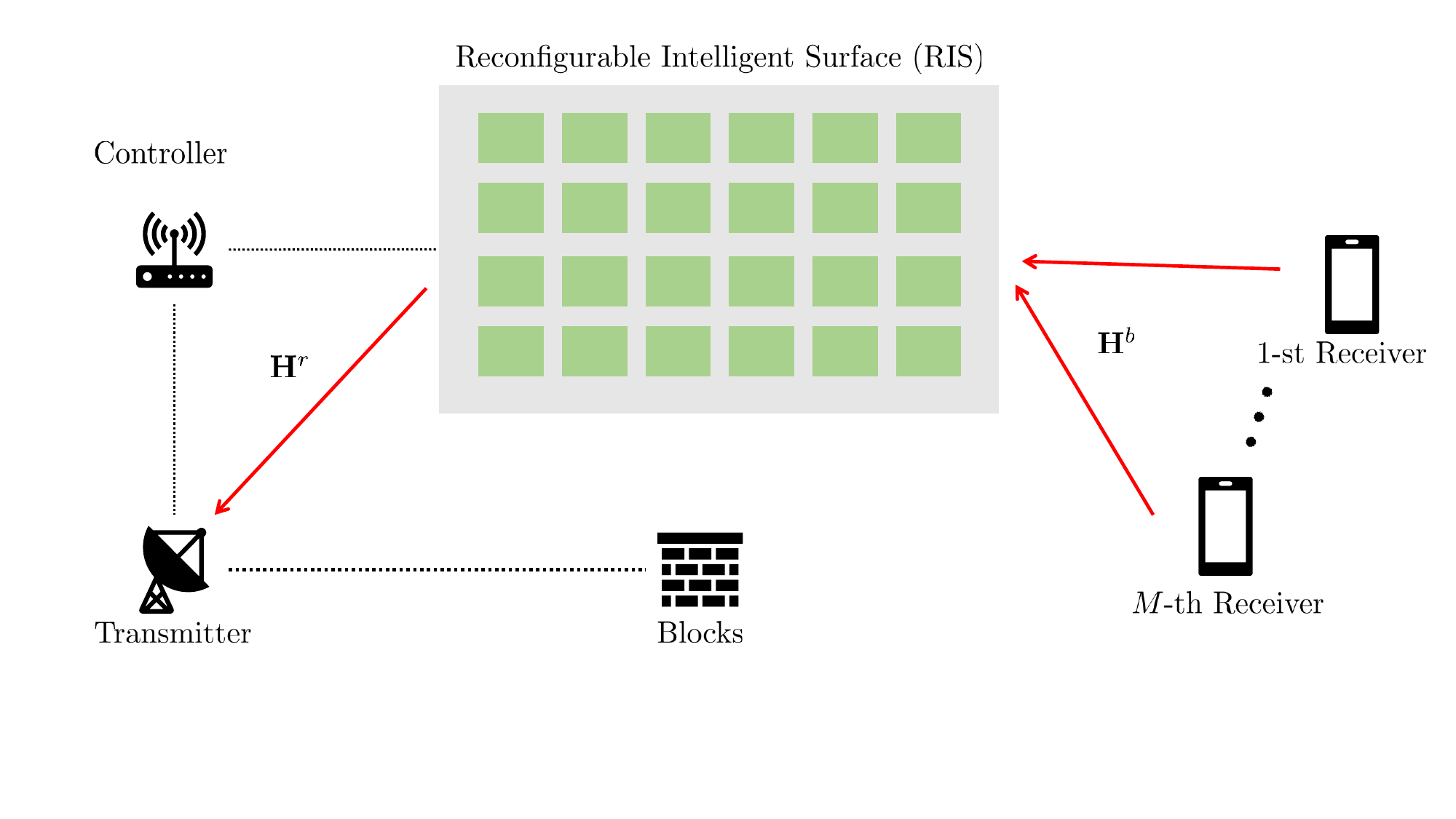}}  \vspace{-0mm}
			\caption{  An uplink RIS-based wireless communication system consisting of a $K$-antenna BS and $M$ single-antenna mobile users. }
			\label{fig:system_model} \vspace{-1cm}
		\end{center}
	\end{figure}

	We consider the  {uplink} communication between a BS equipped with $K$ antenna elements and $M$ single-antenna mobile users, as shown in the Fig.~\ref{fig:system_model}. We assume that this communication is realized via a discrete-element RIS deployed on the facade of a building in the vicinity of the BS side. The RIS is comprised of $N$ unit cells of equal small size, and each made from metamaterials that are capable of adjusting their reflection coefficients. We assume that there is no direct signal path between the BS and the mobile users due to unfavorable propagation conditions, e.g., the presence of large obstacles.  {It should be noted that the existence of direct link has no impact on the proposed algorithms in this work, since direct link can be estimated with the off state of RIS elements using conventional methods.}

	The received discrete-time signals at BS from all $M$ mobile users for $T$ consecutive time slots   can be compactly expressed with $\tilde{\mathbf{Y}} \in\mathbb{C}^{K \times T}$ given by
	\begin{equation}\label{equ:YXZ}
		\setlength\abovedisplayskip{3pt}%shrink space
		\setlength\belowdisplayskip{3pt}
		\tilde{\mathbf{Y}} \triangleq \tilde{\mathbf{H}}^{r}  \mathbf{\Phi}   \tilde{\mathbf{H}}^{b} \tilde{\mathbf{X}}+\tilde{\mathbf{W}},
	\end{equation}
	where diagonal matrix $\mathbf{\Phi}$ is the phase configuration for $N$ RIS unit elements,  {which is full rank and usually chosen from low resolution discrete sets \cite{9110869}}; $\tilde{\mathbf{H}}^{b}\in\mathbb{C}^{N\times M}$ and $\tilde{\mathbf{H}}^{r}\in\mathbb{C}^{K\times N}$ denote the channel matrices between all users and RIS, and between RIS and BS, respectively; the matrix  $\tilde{\mathbf{X}}\in\mathbb{C}^{M \times T}$ includes the $M$ users transmitted signal within $T$ time slots; and $\tilde{\mathbf {W}} \in \mathbb{C}^{K \times T}$ is the Additive White Gaussian Noise (AWGN) matrix with element having zero mean and  variance $N_0$.
    
	 {In typical cellular configuration, the involved narrowband channels are correlated random vectors that are dependent of scattering geometry, however, for large number of antenna elements at RIS, the channels can be represented by sparse matrices in beam domain \cite{9367220,6940305, 837052,7727995,8477183}} Thus, the $n$-th column in channel $\tilde{\mathbf{H}}^{r}$ is expressed as \cite{8322235}
	\begin{equation}
		\setlength\abovedisplayskip{3pt}%shrink space
		\setlength\belowdisplayskip{3pt}
		\begin{aligned}
			\tilde{\mathbf{H}}^{r}_{:,n}=\sum_{l=1}^{N_{c}} \alpha_{n, l} e^{-j 2 \pi d_{n, l} / \lambda_{c}} \mathbf{e}\left(\phi_{n, l}\right),   
		\end{aligned}
	\end{equation}
    where $N_{c}$ is the number of channel paths, $\alpha_{n, l} \sim \mathcal{C} \mathcal{N}(0,1)$ and $d_{n, l}, \phi_{n, l}$ are the channel attenuation, physical distance and AoD associated with the $l$-th path for the $n$-th element, respectively, $\lambda_{c}$ denotes the signal wavelength, $\mathbf{e}\left(\phi_{n, l}\right) \in \mathbb{C}^{K \times 1}$ is the element array vector at the RIS side along with the direction of $\phi_{k, l}$, which satisfies $\left\|\mathbf{e}\left(\phi_{n, l}\right)\right\|_{2}^{2}=$ 1 and can be given by
    	\begin{equation}
    		\setlength\abovedisplayskip{3pt}%shrink space
    		\setlength\belowdisplayskip{3pt} 
    	\mathbf{e}\left(\phi_{n, l}\right)\!=\!\frac{1}{\sqrt{K}}\left[1, e^{j \frac{2 \pi d}{\lambda_{c}} \sin\! \phi_{n, l}}, \!\ldots\!, e^{j \frac{2 \pi d}{\lambda_{c}}(K-1) \sin \!\phi_{n, l}}\right]^{T}\!,
    \end{equation}
    where $d$ is the element spacing. Thus, the channel can be further represented by
    \begin{equation}
    	\setlength\abovedisplayskip{3pt}%shrink space
    	\setlength\belowdisplayskip{3pt}
    	\begin{aligned}
    		\tilde{\mathbf{H}}^{r} = \mathbf{F}_1 \mathbf{H}^{r}
    	\end{aligned}
    \end{equation}
    where $\mathbf{H}^{r}$ is the angular domain channel, and $ \mathbf{F}_1$ is  $K \times K$ discrete Fourier transform (DFT) matrix, the $q$-th column of which is given by 
    \begin{equation}
    	\setlength\abovedisplayskip{3pt}%shrink space
    	\setlength\belowdisplayskip{3pt}
    	[\mathbf{F}_1]_{:, q}=\frac{1}{\sqrt{K}}\left[1, e^{-j \frac{2 \pi q}{K}}, \cdots, e^{-j \frac{2 \pi(K-1) q}{K}}\right]^{T}. 
    \end{equation}

		Thus, we have 
		\begin{equation}
			\setlength\abovedisplayskip{3pt}%shrink space
			\setlength\belowdisplayskip{3pt}
			\begin{aligned}
				\mathbf{H}^{r}&=\mathbf{F}_1 ^H  \tilde{\mathbf{H}}^{r} = \mathbf{F}_1 ^H \mathbf{E} \boldsymbol{\beta}, 
			\end{aligned}
		\end{equation}
		where  $\mathbf{E}_n =\left[\mathbf{e}\left(\phi_{n, 1}\right), \mathbf{e}\left(\phi_{n, 2}\right), \cdots, \mathbf{e}\left(\phi_{n, N_{c}}\right)\right] \in \mathbb{C}^{K \times N_{c}}$ and $\boldsymbol{\beta}_n =\left[\beta_{n, 1}, \beta_{n, 2}, \cdots, \beta_{ n, N_{c}}\right]^{T} \in \mathbb{C}^{N_{c} \times 1}$ with $\beta_{n, l}=$
		$\alpha_{n, l} e^{-j 2 \pi d_{n, l} / \lambda_{c}}$, and $\mathbf{E}\boldsymbol{\beta}$ is the collection of  $\mathbf{E}_n \boldsymbol{\beta}_n$.  Based on \cite{8322235}, the equivalent channel $\mathbf{H}^{r}$ is sparse. 
		
		Similarly, the channel  $\tilde{\mathbf{H}}^{b}$ can also be represented by sparse beam domain channel
		\begin{equation}
			\setlength\abovedisplayskip{3pt}%shrink space
			\setlength\belowdisplayskip{3pt}
			\begin{aligned}
				\tilde{\mathbf{H}}^b= \mathbf{H}^b \mathbf{F}_2,
			\end{aligned}
		\end{equation}
		where the precoding matrix  $\mathbf{F}_2$ is the $M \times M$ DFT matrix.

    Thus, the input-output relationship \eqref{equ:YXZ} can be rewritten as 
	\begin{equation}
	\tilde{\mathbf{Y}} \triangleq  \mathbf{F}_1^H {\mathbf{H}}^{r}  \mathbf{\Phi}   {\mathbf{H}}^{b} \mathbf{F}_2^H \tilde{\mathbf{X}}+\tilde{\mathbf{W}} \Rightarrow \mathbf{Y}\triangleq\mathbf{H}^{r}  \mathbf{\Phi}  \mathbf{H}^{b} \mathbf{X} + \mathbf{W}, 
	\end{equation}
	where  $\mathbf{Y}=\mathbf{F}_1 \tilde{\mathbf{Y}}$, $\mathbf{X}=\mathbf{F}_2^H \tilde{\mathbf{X}}$ and $ \mathbf{W}=\mathbf{F}_1 \tilde{\mathbf{W}}$. Therefore, the beam domain representation yields an equivalent sparse channel estimation in beamspace. Generally, we can obtain the locally optimal performance or approximately optimal performance in separate channel estimation or signal recovery at the price of heavy training and computing. To cascade these locally  optimal channel estimation and signal recovery together,  the cascaded result usually is not the optimal performance for the whole communication system. To address this problem, we initially implement the joint channel  estimation and signal recovery scheme without consuming too much computational complexity. Specifically,  our objective is to jointly estimate the channel ${\mathbf{H}}^r$ and ${\mathbf{H}}^b$,  and transmitted signal $\mathbf{X}$ during the signal transmission phase, which is provided in the following section.  
	
	\section{Problem Formulation and Message Passing Algorithm}\label{sec:channel_est}
	\subsection{Problem Formulation and Factor Graph Representation}
\begin{figure}[h]\vspace{-0mm}
		\begin{center}
			\centerline{\includegraphics[width=0.5\textwidth]{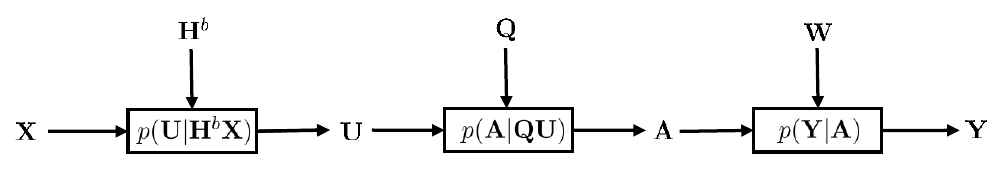}}  \vspace{-0mm}
			\caption{The framework of the formulated problem.} 
			\label{fig:multi_layer}  
		\end{center}
	\end{figure}

	The focus of this paper is to design an efficient receiver to estimate transmitted signal $\mathbf{X}$ and all involved channels $\mathbf{H}^b$ and $\mathbf{H}^r$. To this end, we formulate a two-layer estimation problem, and the framework is shown in Fig.~\ref{fig:multi_layer}. In the first layer, the input is transmitted signal $\mathbf{X}$, and the output is $\mathbf{U}=\mathbf{H}^b \mathbf{X}$. In the second layer, the input is $\mathbf{U}$, and the output is $\mathbf{A}=\mathbf{Q}  \mathbf{U}$, with $\mathbf{Q}=\mathbf{H}^r \mathbf{\Phi}$ that incorporates the unknown channel $\mathbf{H}^r$,  {and $\mathbf{\Phi}$ is assumed to be known}. The output $\mathbf{A}$ is corrupted by the noise $\mathbf{W}$, which is interpreted as $\mathbf{Y}=\mathbf{A}+\mathbf{W}$. 
	\vspace{-1cm}
    
    Thus, the joint probability $p(\mathbf{Q},  \mathbf{H}^b,  \mathbf{X}, \mathbf{Y}) $ can be factorized into 
    \begin{equation} \label{equ:factor}
    	\setlength\abovedisplayskip{3pt}%shrink space
    	\setlength\belowdisplayskip{3pt} 
    	\begin{aligned}
    		& p\!(\!\mathbf{Q}\!, \! \mathbf{H}^b\!, \! \mathbf{X}, \!\!\! \mathbf{Y}\!)\! \! \propto\! \!p(\!\mathbf{X}\!) p( \!\mathbf{H}^b \!) p(\!\mathbf{U} \!\!\mid\! \!\mathbf{H}^b\! \mathbf{X}\!) p(\! \mathbf{Q}\!)  p(\!\mathbf{A} \!\!\mid \!\! \mathbf{Q}\! \mathbf{U}\!) p(\!\mathbf{Y}\! \!\mid\!\! \mathbf{A}\!)\!.
    	\end{aligned}
    \end{equation}

    \begin{figure} 
    	\begin{center}
    		\centerline{\includegraphics[width=0.5\textwidth]{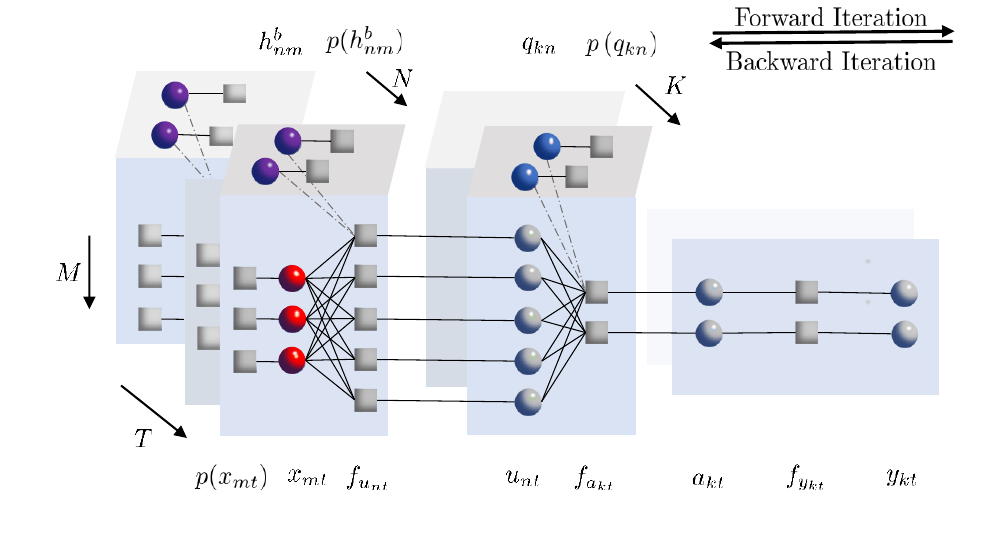} } \vspace{-2mm}
    		\caption{The factor graph of the joint channel estimation and signal recovery in RIS-empowered wireless communication systems.}
    		\label{fig:factor_graph} 
    	\end{center}
    \end{figure} 

    The probabilistic structure characterized by \eqref{equ:factor} is illustrated by Fig.~\ref{fig:factor_graph}. The circles represent variables and the squares represent factors. The purple circles denote the variable $h_{nm}^b$; the red circles are variables $x_{mt}$; and the blue circles represent $q_{kn}$. $p(x_{mt}), p(h_{nm}^b)$ and $p(q_{kn})$ are Gaussian priors of variables $x_{mt}, h_{nm}^b$ and $q_{kn}$, respectively. $f_{u_{nt}}$ is the $(n,t)$-th entry of $p(\mathbf{U} \mid \mathbf{H}^b \mathbf{X})$; $f_{a_{kt}}$ is the $(k,t)$-th entry of $ p(\mathbf{A} \mid  \mathbf{Q} \mathbf{U})$; and $f_{y_{kt}}$ is the $(k,t)$-th element of $p(\mathbf{Y} \mid \mathbf{A})$. As shown in Fig.~\ref{fig:factor_graph}, the message updates bidirectionally, where the message flows from the left to right is termed as forward iteration, and the message flows conversely is backward iteration. The definitions of involved messages are summarized in Table.~\ref{table:messages}, and according to the sum-product message passing rules, meassage updates in each layer are presented as follows:
 
	\begin{table}
		\caption{Message definitions in the factor graph.}
		\centering
		\scalebox{0.8 }{
			\begin{tabular}{|m{1.3cm}|m{3.2cm}|m{1.3cm}|m{3.2cm}|}	\hline 
				\multicolumn{4}{|c|}{\textbf{The first layer}} \\ \hline
			 $\mu_{x_{mt} \leftarrow f_{u_{nt}}}  $& message from $f_{u_{nt}}$ to $x_{mt}$  &$	\mu_{x_{mt} \rightarrow f_{u_{nt}}}$ &message from $x_{mt}$ to $f_{u_{nt}}$\\
				\hline   
				$\mu_{f_{u_{nt}} \leftarrow h_{nm}^b}$&message from $h_{nm}^b$ to $f_{u_{nt}}$ &$\mu_{f_{u_{nt}} \rightarrow h_{nm}^b}$& message from $f_{u_{nt}}$ to $h_{nm}^b$  \\
				\hline
				$\mu_{f_{u_{nt}} \leftarrow u_{nt} } $ & message from $u_{nt}$ to $f_{u_{nt}}$& $\mu_{f_{u_{nt}} \rightarrow u_{nt} }$ & message from $f_{u_{nt}}$ to $u_{nt}$ \\
				\hline
				\multicolumn{4}{|c|}{\textbf{The second layer}} \\ \hline
			   $\mu_{u_{nt} \leftarrow f_{a_{kt}} }$ &  message from $f_{a_{kt}}$ to $u_{nt}$ &$\mu _{u_{nt} \rightarrow f_{a_{kt}}} $ & message from $u_{nt}$ to $f_{a_{kt}}$ \\
				\hline
		       	 $\mu _{f_{a_{kt}} \rightarrow q_{kn}}$ &  message from $f_{a_{kt}}$ to $q_{kn}$ &  $\mu_{f_{a_{kt}} \leftarrow q_{kn}} $ & message from $q_{kn}$ to $f_{a_{kt}}$ \\
				\hline
				  $\mu _{f_{a_{kt}} \rightarrow a_{kt} }$ &  message from $f_{a_{kt}}$ to $a_{kt}$ & $\mu _{f_{a_{kt}} \leftarrow a_{kt} }$ &  message from $a_{kt}$ to $f_{a_{kt}}$ \\
				\hline
				 $	\mu_{a_{kt}  \rightarrow  f_{y_{kt}} }$ &  message from $a_{kt}$ to $f_{y_{kt}}$  & $\mu_{f_{y_{kt}}  \rightarrow  y_{kt} }$ &  message from $f_{y_{kt}}$ to $y_{kt}$ \\
				\hline
				 $\mu_{a_{kt}  \leftarrow  f_{y_{kt}} }$ &  message from $f_{y_{kt}}$ to $a_{kt}$& & \\
				\hline
		\end{tabular}}
		\label{table:messages}
	\end{table}
 
	\subsubsection{The First Layer}
	The message from $f_{u_{nt}}$ to $x_{mt}$ in the $\ell$-th iteration is given by \vspace{-0.15cm}
	\begin{equation}
		\setlength\abovedisplayskip{3pt}%shrink space
		\setlength\belowdisplayskip{3pt} 
	    \begin{aligned}
	    	&\mu^{\ell}_{x_{mt} \leftarrow f_{u_{nt}}} (x_{mt})  \\
	    	&=  f_{u_{nt}}^{\ell}  \mu_{f_{u_{nt}} \leftarrow  u_{nt}}^{\ell} \prod_{m'\neq m}^{M} {\mu_{x_{m' t} \rightarrow f_{u_{nt}}}^{\ell}} \prod_{m=1}^{M} {\mu_{f_{u_{nt}} \leftarrow  h^b_{nm} }^{\ell}}.
	    \end{aligned}
	\end{equation} 
	
	The message from $x_{mt}$ to $f_{u_{nt}}$  in the $(\ell+1)$-th iteration  is 
	\begin{equation}
		\setlength\abovedisplayskip{3pt}%shrink space
		\setlength\belowdisplayskip{3pt}
		\mu^{\ell+1}_{x_{mt} \rightarrow f_{u_{nt}}} (x_{mt}) = p(x_{mt}) \prod_{n' \neq n}^{N} {\mu^{\ell}_{x_{mt} \leftarrow f_{u_{n't}}} (x_{mt})}. 
	\end{equation}
	
	The message from $f_{u_{nt}}$ to $h_{nm}^b$ in the $\ell$-th iteration can be expressed as 
	\begin{equation}
		\setlength\abovedisplayskip{3pt}%shrink space
		\setlength\belowdisplayskip{3pt} 
		\begin{aligned}
			&\mu^{\ell}_{f_{u_{nt}} \rightarrow h_{nm}^b} (h_{nm}^b) \\
			&=  f_{u_{nt}}^{\ell} \mu_{f_{u_{nt}} \leftarrow  u_{nt}}^{\ell}  \prod_{m'\neq m}^{M} {\mu_{f_{u_{nt}} \leftarrow h_{nm'}^b}^{\ell}}  \prod_{m=1}^{M} \mu^{\ell}_{x_{mt}  \rightarrow {f_{u_{nt}}}}.
		\end{aligned}
	\end{equation}
	
	The message from $h_{nm}^b$ to $f_{u_{nt}}$  in the $(\ell+1)$-th iteration  is 
	\begin{equation}
		\setlength\abovedisplayskip{3pt}%shrink space
		\setlength\belowdisplayskip{3pt} 
		\mu^{\ell+1}_{f_{u_{nt}} \leftarrow h_{nm}^b} ( h_{nm}^b) = p( h_{nm}^b) \prod_{t' \neq t}^{T} {\mu^{\ell}_{ f_{u_{nt'}} \rightarrow h_{nm}^b} } (h_{nm}^b).
	\end{equation}
	
	The message from $u_{nt}$ to $f_{u_{nt}}$ in the $\ell$-th iteration is given by
	\begin{equation}
		\setlength\abovedisplayskip{3pt}%shrink space
		\setlength\belowdisplayskip{3pt} 
		\mu^{\ell}_{f_{u_{nt}} \leftarrow u_{nt} } (u_{nt}) = \prod_{k=1}^{K} {\mu^{\ell}_{u_{nt} \leftarrow f_{a_{kt} } } }.
	\end{equation}
	
	The message from $f_{u_{nt}}$ to $u_{nt}$ in the $\ell$-th iteration is 
	\begin{equation}
		\setlength\abovedisplayskip{3pt}%shrink space
		\setlength\belowdisplayskip{3pt} 
		\mu^{\ell}_{f_{u_{nt}} \rightarrow u_{nt} } (u_{nt}) = f_{u_{nt}}  \prod_{m=1}^{M} {\mu^{\ell}_{x_{mt} \rightarrow f_{u_{nt} } } }  \!\! \prod_{m=1}^{M} \!\!{\mu^{\ell}_{f_{u_{nt} } \leftarrow  h_{nm}^b} }.
	\end{equation}

	\subsubsection{The Second Layer}
	The message from $f_{a_{kt}}$ to $u_{nt}$ in the $\ell$-th iteration is 
	\begin{equation}
		\setlength\abovedisplayskip{3pt}%shrink space
		\setlength\belowdisplayskip{3pt} 
	\begin{aligned}
		&\mu^{\ell}_{u_{nt} \leftarrow f_{a_{kt}} } (u_{nt} ) \\
		&=  f_{a_{kt}}^{\ell} \mu^{\ell}_{f_{a_{kt}} \leftarrow a_{kt}  } \prod_{n' \neq n}^{N} {\mu_{ u_{n't} \rightarrow  f_{a_{kt}}}^{\ell} }   \prod_{n=1}^{N} {\mu^{\ell}_{f_{a_{kt}}  \leftarrow  q_{kn}} }.
	\end{aligned}
	\end{equation}
	
	The message from $u_{nt}$ to $f_{a_{kt}}$   in the $\ell$-th iteration is 
	\begin{equation}
		\setlength\abovedisplayskip{3pt}%shrink space
		\setlength\belowdisplayskip{3pt} 
		\mu^{\ell+1}_{u_{nt} \rightarrow f_{a_{kt}}} (u_{nt}) = \mu_{ f_{u_{nt}} \rightarrow u_{nt} } \prod_{k' \neq k}^{K} {\mu^{\ell}_{u_{nt} \leftarrow f_{z_{k't}} } (u_{nt})}.
	\end{equation}

	The message from $f_{a_{kt}}$ to $q_{kn}$  in the $\ell$-th iteration  is 
	\begin{equation}
		\setlength\abovedisplayskip{3pt}%shrink space
		\setlength\belowdisplayskip{3pt} 
		\begin{aligned}
			&\mu^{\ell}_{f_{a_{kt}} \rightarrow q_{kn}} (q_{kn}) \\
			&=  f_{a_{kt}}^{\ell} \mu_{f_{a_{kt}} \leftarrow  a_{kt}}^{\ell}  \prod_{n'\neq n}^{N} {\mu_{f_{a_{kt}} \leftarrow q_{kn'}}^{\ell}}  \prod_{n=1}^{N} \mu^{\ell}_{u_{nt}  \rightarrow {f_{a_{kt}}}}.
		\end{aligned}
	\end{equation}

	The message from $q_{kn}$ to $f_{a_{kt}}$  in the $(\ell+1)$-th iteration  is 
	
	\begin{equation}
		\setlength\abovedisplayskip{3pt}%shrink space
		\setlength\belowdisplayskip{3pt} 
		\mu^{\ell+1}_{f_{a_{kt}} \leftarrow q_{kn}} ( q_{kn}) = p( q_{kn} ) \prod_{t' \neq t}^{T} {\mu^{\ell}_{ f_{v_{kt'}} \rightarrow q_{kn}} } (q_{kn}).
	\end{equation}

	\subsubsection{The Output Layer}
	The message from $f_{a_{kt}}$ to $a_{kt}$ in the $\ell$-th iteration is  \vspace{-0.3cm}
	\begin{equation}  
		\mu^{\ell}_{f_{a_{kt}} \rightarrow a_{kt} } (a_{kt}) = f_{a_{kt}}  \prod_{n=1}^{N} {\mu^{\ell}_{  f_{a_{kt} \leftarrow q_{kn}  } } }   \prod_{n=1}^{N} {\mu^{\ell}_{ u_{nt}\rightarrow  f_{a_{kt} }} }.
	\end{equation}
	
	The message from $a_{kt}$ to $f_{a_{kt}}$ in the $\ell$-th iteration is 
	\begin{equation}
		\mu^{\ell}_{f_{a_{kt}} \leftarrow a_{kt} } (a_{kt}) = {\mu^{\ell}_{a_{kt} \leftarrow f_{y_{kt} } } (y_{kt})}.
	\end{equation}
	
	The message from $a_{kt}$ to $f_{y_{kt}}$ is
	\begin{equation}
		\mu_{a_{kt}  \rightarrow  f_{y_{kt}} }^{\ell} (y_{kt}) = \mu^{\ell}_{f_{a_{kt}} \rightarrow a_{kt} } (a_{kt}),
	\end{equation}
	
	and the message from $f_{y_{kt}}$ to $y_{kt}$ equals to 
	\begin{equation}
		\mu_{f_{y_{kt}}  \rightarrow  y_{kt} } (y_{kt}) = f_{y_{kt}} \mu^{\ell}_{a_{kt} \rightarrow f_{y_{kt}} } (a_{kt}). 
	\end{equation}
	
	The message from $f_{y_{kt}}$ to $a_{kt}$ is  
	\begin{equation}
		\mu_{a_{kt}  \leftarrow  f_{y_{kt}} } (a_{kt}) = f_{y_{kt}} \mu^{\ell}_{f_{y_{kt}} \leftarrow y_{kt} } (y_{kt}).
	\end{equation}

	\subsubsection{Beliefs of Estimated Variables}
	The beliefs (approximate a posterior distributions) of $x_{mt}$, $h_{nm}^b$ and $q_{kn}$ are respectively given by\vspace{-0.5cm}
	\begin{equation}
		\setlength\abovedisplayskip{3pt}%shrink space
		\setlength\belowdisplayskip{3pt} 
		\mu_{x_{mt}}  (x_{mt}) = p({x_{mt}}) \prod_{n=1}^{N} \mu_{ x_{mt} \leftarrow f_{u_{nt}}  }  (x_{mt}),
	\end{equation}	
\vspace{-0.15cm}
	\begin{equation}
		\setlength\abovedisplayskip{3pt}%shrink space
		\setlength\belowdisplayskip{3pt} 
		\mu_{h_{nm}^b}  (h_{nm}^b) = p(h_{nm}^b) \prod_{t=1}^{T} \mu_{ f_{u_{nt}}  \rightarrow h_{nm}^b }  (h_{nm}^b),
	\end{equation}	\vspace{-0.2cm}
	\begin{equation}
		\setlength\abovedisplayskip{3pt}%shrink space
		\setlength\belowdisplayskip{3pt} 
		\mu_{q_{kn}}  (q_{kn}) = p(q_{kn}) \prod_{t=1}^{T} \mu_{ f_{a_{kt}}  \rightarrow q_{kn} }  (q_{kn}).
	\end{equation}
\vspace{-0.15cm}
	
	%% Approximated messages
	\subsection{Derivation and Approximation of Message Passing} \label{sec:joint_ce_sr}
	We have derived the sum product algorithm with the factor graph in Fig.~\ref{fig:factor_graph}, however, exact implementation is impractical due to the involvement of numerous loops and both discrete and continuous-valued variables.  Thus, the Gaussian approximation and Taylor series expansion are applied to further simplify the messages of loopy belief propagation (LBP) for efficient inference. In this section, we derive the approximate message passing algorithms for joint channel estimation and signal recovery in Fig.~\ref{fig:factor_graph}.  {The detailed derivation of each message is provided in Appendix. Here, we give the simplified messages processed by Gaussian approximated and Taylor series expansion.}
		
	 {In the first layer, the message from factor nodes to variables can be simplified as} 
   \begin{equation}
   	\setlength\abovedisplayskip{3pt}%shrink space
   	\setlength\belowdisplayskip{3pt} 
   	\begin{aligned}
   		&\mu_{x_{mt}\! \leftarrow \!f_{u_{nt}}}\! (\!x_{mt}\!)\! \sim \!\mathcal{N} \!\!\left( \! \! x_{mt} \!\mid \!\frac {\hat{h}_{nm}^b \tilde{s}_{nt} +  \hat{x}_{mt} |\hat{h}_{nm}^b|^2  v^s_{nt}} {|\hat{h}_{nm}^b|^2  v^s_{nt} + v^b_{nm} v^s_{nt} - v^b_{nm} |\tilde{s}_{nt}|^2 } , \right. \\
   		&\qquad \qquad \qquad \left. \frac{1} {|\hat{h}_{nm}^b|^2  v^s_{nt} + v^b_{nm} v^s_{nt} - v^b_{nm} |\tilde{s}_{nt}|^2 }  \right),
   	\end{aligned}
   \end{equation}
\begin{equation}\notag
	\setlength\abovedisplayskip{3pt}%shrink space
	\setlength\belowdisplayskip{3pt} 
	\begin{aligned}
   		& \mu_{f_{u_{nt}}\!\! \rightarrow \!h_{nm}^b}\! (h_{nm}^b)\!  \sim \!\mathcal{N}\!\! \left(\! h_{nm}^b \!\mid\!  \frac {\hat{x}_{mt}  \tilde{s}_{nt} +  \hat{h}_{nm}^b |\hat{x}_{mt}^b|^2  v^s_{nt}} {|\hat{x}_{mt}|^2  v^s_{nt} + v^x_{mt} v^s_{nt} - v^x_{mt} |\tilde{s}_{nt}|^2 } ,  \right. \\
   		&\qquad \qquad\qquad \left. \frac{1} {|\hat{x}_{mt}|^2  v^s_{nt} + v^x_{mt} v^s_{nt} - v^x_{mt} |\tilde{s}_{nt}|^2 }  \right).
   	\end{aligned}
   \end{equation}
    
     {The beliefs of two variables $x_{mt}$ and $h_{nm}^b$ are given as}
    \begin{equation}
    \begin{aligned}
    	&\mu_{x_{mt} } (x_{mt}) \sim   \mathcal{N} \left( x_{mt } \mid \hat x_{mt  }, v^x_{mt }  \right), \\ 
    	&\mu_{h_{nm}^b}  (h_{nm}^b)  \sim \mathcal{N} \left( h_{nm}^b \mid \hat{h}_{nm}^b, v^b_{nm} \right).
    \end{aligned}
	\end{equation}
    
     {In the second layer, the message from factor nodes to variables can be simplified as}
    \begin{equation}
    	\setlength\abovedisplayskip{3pt}%shrink space
    	\setlength\belowdisplayskip{3pt} 
    	\begin{aligned}
    		&\mu _{u_{nt}\! \leftarrow \! f_{v_{kt}}}\!(u_{nt}) \! \!\sim\! \mathcal{N}\!\! \left( \!\!  u_{nt}\! \!\mid\! \frac {\hat{q}_{kn} \tilde{s}_{kt} +  \hat{u}_{nt} |\hat{q}_{kn}|^2  v^s_{kt}} {|\hat{q}_{kn}|^2  v^s_{kt} + v^q_{kn} v^s_{kt} - v^q_{kn} |\tilde{s}_{nt}|^2 } ,  \right. \\
    		&\qquad \qquad\qquad \left. \frac{1} {|\hat{q}_{kn}|^2  v^s_{kt} + v^q_{kn} v^s_{kt} - v^q_{kn} |\tilde{s}_{kt}|^2 }  \right), \\
    		&\mu_{f_{a_{kt}} \!\rightarrow\! q_{kn}} \!(q_{kn})\! \!\sim\! \mathcal{N} \!\!\left( \!\!  q_{kn}\! \!\mid\! \frac {\hat{u}_{nt} \tilde{s}_{kt} +  \hat{q}_{kn} |\hat{u}_{nt}|^2  v^s_{kt}} {|\hat{u}_{nt}|^2  v^s_{kt} + v^u_{nt} v^s_{kt} - v^u_{nt} |\tilde{s}_{nt}|^2 } ,  \right. \\
    		&\qquad \qquad \qquad\left. \frac{1} {|\hat{u}_{nt}|^2  v^s_{kt} + v^u_{nt} v^s_{kt} - v^u_{nt} |\tilde{s}_{kt}|^2 }  \right).
    	\end{aligned}
    \end{equation}
    
     {The belief of variable $q_{kn}$ is}
    \begin{equation}  
    	\setlength\abovedisplayskip{3pt}%shrink space
    	\setlength\belowdisplayskip{3pt} 
    	\mu_{q_{kn}}  (q_{kn})  \sim \mathcal{N} \left( q_{kn} \mid \hat{q}_{kn}, v^q_{kn} \right).
    \end{equation}

	\subsection{Two Proposed Schemes} \label{sec:algorithms}
The proposed bidirectional channel estimation and signal recovery problem is given by the factor graph Fig.~\ref{fig:factor_graph}, which is mainly composed of backward part and forward part. The backward part firstly estimates the intermediate variables of matrix $\mathbf{Q}$  and  matrix $\mathbf{U}$ with the aid of received signals in the second layer, then the intermediate variables of signal $\mathbf{X}$ and $\mathbf{H}^b$ are estimated  in the first layer, as shown in Alg.~\ref{alg:als_back}. In the forward part, the signal $\mathbf{X}$ and $\mathbf{H}^b$ are estimated in the first layer, then the unknown matrix $\mathbf{Q}$ is estimated in the second layer, as shown in  Alg.~\ref{alg:als_forw}. The overall algorithm is shown in Alg.~\ref{alg:als_back} and Alg.~\ref{alg:als_forw}.  {The parameter $\epsilon$ is the threshold in the  stop criterion, and it is selected as a small number, which is normally less than or equal to $10^{-4}$.} To further improve the estimation performance, we adopt an inner iteration in the estimation of matrix $\mathbf{Q}$, which leads to two schemes, BAMP algorithm and BUTAMP two-layer algorithm.  
 
	\subsubsection{BAMP Two-layer Algorithm}
	We propose a BAMP algorithm to refine the estimation of matrix $\mathbf{Q}$,  which improves the convergence of the whole algorithm. Specifically, the pilots $\mathbf{X}^p \in \mathbb{R}^{M\times T_p}$ are used in the initial iteration, and estimates of $\mathbf{X}$ and $\mathbf{H}^b$ are obtained in the first layer, then, the output $\hat{\mathbf{U}}=\hat{\mathbf{H}}^b  \hat{\mathbf{X}}$ is considered as input of the estimation of $\mathbf{Q}$. Inspired by the GAMP algorithm, the obtained $\mathbf{Q}$ is applied in subsequent iterations. Although the BAMP two-layer algorithm can estimate all involved unknown channels and signal simultaneously, the divergency issue still exists. This arises from the ill-conditioned matrix $\mathbf{U}$, which is the product of two Gaussian distributed matrices $\mathbf{X}$ and $\mathbf{H}^b$. To improve the convergence, a damping factor $\beta$ is introduced to the intermediate variables $v^s$ in step 5 and 13 of Alg.~\ref{alg:als_back} and $V$ in step 7 and 15 in Alg.~\ref{alg:als_forw}, i.e., ${v^s}^{(\ell)}=(1-\beta){v^s}^{(\ell-1)}+\beta {v^s}^{(\ell)}$ and  ${V}^{(\ell)}=(1-\beta){V}^{(\ell-1)}+\beta {V}^{(\ell)}$.
	 \subsubsection{BUTAMP Two-layer Algorithm}
	 BAMP is sensitive to the damping factor, thus, to further improve the convergence, the BUTAMP is proposed for strong robustness. Specifically, the estimates $\mathbf{X}$ and $\mathbf{H}^b$ are obtained in the first layer with the aid of pilots, then, the singular value decomposition (SVD) is performed for the estimate  $\mathbf{U}=\boldsymbol{\mathcal{U}}\mathbf{\Lambda}\boldsymbol{\mathcal{V}}$. Thus, a new input-output relationship can be given as 
	 \begin{equation}\label{equ:utamp}
	 	\setlength\abovedisplayskip{3pt}%shrink space
	 	\setlength\belowdisplayskip{3pt}
	 	\begin{aligned}
	 		  \mathbf{Y}=\mathbf{Q} \mathbf{U} + \mathbf{W} &\Rightarrow \mathbf{Y}\boldsymbol{\mathcal{V}}^H=\mathbf{Q} \boldsymbol{\mathcal{U}}\mathbf{\Lambda} + \mathbf{W} \boldsymbol{\mathcal{V}}^H \\
	 		  & \Rightarrow \mathbf{R}=\mathbf{Q} \tilde{\mathbf{U}} + \tilde{\mathbf{W}},
	 	\end{aligned}
	 \end{equation}
     where $\tilde{\mathbf{U}}=\boldsymbol{\mathcal{U}}\mathbf{\Lambda}$ is the transformed measurement matrix and $\mathbf{R}=\mathbf{Y}\boldsymbol{\mathcal{V}}^H$ is the transformed received signal.
	  Then, the estimate of  $\mathbf{Q}$ can be obtained from \eqref{equ:utamp} using  AMP with unitary transformation \cite{9293406} with low computational cost.  
 \vspace{-0.3cm}
	
	\begin{algorithm}[h]\caption{\textbf{Joint CE and Signal Detection Algorithm (1): Backward Iteration}}
		\begin{algorithmic}[1]
			\Require The received signal $y_{kt}$, and the number of maximum algorithmic iterations $L$. 
			\textbf{Initialization:} Initialize $x_{mt}=0,v_{mt}^x=1,  h_{nm}^b=0, v_{nm}^b=1, a_{kt}=0$, and $v^a_{kt}=1$.
			
			\State \textbf{for} {$\ell=1,2,\ldots,L$} \textbf{do}
			\State  \quad  $\tilde v_{kt}^{(2)}=\left(\frac{1}{V_{kt}^{(2)} }+\frac{1}{v^{w}}\right)^{-1}$
			\State  \quad $\tilde z_{kt}^{(2)}=\tilde v_{kt}^{(2)} \left(\frac{Z_{kt}^{(2)}}{V_{kt}^{(2)} }+\frac{y_{kt}}{v^{w}}\right)  $
			\State  \quad $\tilde{s}_{kt}^{(2)} = \frac{   \tilde{z}_{kt}^{(2)} - Z_{kt}^{(2)}  }   {V_{kt}^{(2)}} $
			\State \quad ${v^s_{kt}} ^{(2)} = -\frac{  \tilde{v}_{kt}^{(2)} - V_{kt}^{(2)}  } {{V_{kt}^{(2)}}^2}$
			\State \quad  $\Sigma^{u}_{nt}  =  \left( { \sum_{k=1}^{K}  |\hat{q}_{kn}|^2  {v^s_{kt}}^{(2)}  } \right) ^{-1}$
			\State \quad $R^{u}_{nt}  = \hat{u}_{nt} \left( 1+ \Sigma^{u}_{nt} \sum_{k=1}^{K} {v^q_{kn} |\tilde{s}_{kt}^{(2)}|^2 - v^q_{kn} {v^s_{kt}}^{(2)}}  \right) + \Sigma^{u}_{nt}  \sum_{k=1}^{K} {\hat{q}_{kn} \tilde{s}_{kt}^{(2)}}  $
			\State \quad $\Sigma^{q}_{kn} =  \left( { \sum_{t=1}^{T}  |\hat{u}_{nt}|^2  {v^s_{kt}}^{(2)}  } \right) ^{-1}$
			\State \quad $R^{q}_{kn}	=\hat{q}_{kn} \left( 1+ \Sigma^{q}_{kn} \sum_{t=1}^{T} {v^u_{nt} |\tilde{s}_{kt}^{(2)}|^2 - v^u_{nt} {v^s_{kt}}^{(2)}}  \right) + \Sigma^{q}_{kn}  \sum_{t=1}^{T} {\hat{u}_{nt} \tilde{s}_{kt}^{(2)}}  $
			\Statex \% Message passing in the first layer
			\State  \quad $\tilde{v}_{nt}^{(1)}  =  \left( \frac{1}{V_{nt}^{(1)} } + \frac{1}{\Sigma^{u}_{nt}} \right) ^{-1}$
			\State  \quad $\tilde{z}_{nt}^{(1)} = \tilde{v}^{(1)} _{nt}  \left( \frac{Z_{nt}^{(1)} }{V_{nt}^{(1)} } + \frac{R^{u}_{nt}}{\Sigma^{u}_{nt}} \right)$
			\State  \quad $\tilde{s}_{nt}^{(1)} = \frac{   \tilde{z}_{nt}^{(1)} - Z_{nt}^{(1)}  }   {V_{nt}^{(1)}} $
			\State  \quad ${v^s_{nt}} ^{(1)} = - \frac{  \tilde{v}_{nt}^{(1)} - V_{nt}^{(1)}  } {{V_{nt}^{(1)}}^2}$
			\State  \quad $\Sigma^{x}_{mt}  =  \left( { \sum_{n=1}^{N}  |\hat{h}_{nm}^b|^2  {v^s_{nt}} ^{(1)}  } \right) ^{-1}$
			\State  \quad $R^{x}_{mt} =\hat{x}_{mt} \left( 1- \Sigma^{x}_{mt} \sum_{n=1}^{N} { v^b_{nm} {v^s_{nt}} ^{(1)}}  \right) + \Sigma^{x}_{mt}  \sum_{n=1}^{N} {\hat{h}_{nm}^b \tilde{s}_{nt}^{(1)}}$
			\State \quad  $\Sigma^{b}_{nm} =  \left( { \sum_{t=1}^{T}  |\hat{x}_{mt}|^2  {v^s_{nt}}^{(1)}  } \right) ^{-1}$
			\State  \quad $R^{b}_{nm} =\hat{b}_{nm} \left( 1+ \Sigma^{b}_{nm} \sum_{t=1}^{T} {v^x_{mt} |\tilde{s}_{nt}^{(1)}|^2 - v^x_{mt} {v^s_{nt}} ^{(1)}}  \right) + \Sigma^{b}_{nm}  \sum_{t=1}^{T} {\hat{x}_{mt} \tilde{s}_{nt}^{(1)}}   $			
		\end{algorithmic} \label{alg:als_back}
	\end{algorithm}
\vspace{-0.4cm}

	\begin{algorithm}[h] \caption{\textbf{Joint CE and Signal Detection Algorithm (2): Forward Iteration}}		
		\begin{algorithmic}[1]
			\State \quad $ v^x_{mt} = \left( \frac{1}{\Sigma^{x}_{mt}} +  \frac{1}{v^0_x} \right)^{-1}$
			\State \quad $\hat x_{mt} = v^x_{mt} \left( \frac{R^{x}_{mt}}{\Sigma^{x}_{mt}} +  \frac{\hat x^0}{v^0_x} \right)$
			\State \quad $ v^b_{nm} = \left( \frac{1}{\Sigma^{b}_{nm}} +  \frac{1}{v^0_b} \right)^{-1}$
			\State \quad $\hat h_{nm}^b = v^b_{nm} \left( \frac{R^{b}_{nm}}{\Sigma^{b}_{nm}} +  \frac{\hat h_{0}^b}{v^0_b} \right)$
			\State \quad $\bar V_{nt}^{(1),\ell+1} = \sum_{m=1}^{M}  | \hat{x}_{ mt}^{ \ell+1}|^2 v_{nm}^{b,\ell+1}
			+ |\hat{h}_{ nm}^{ b,\ell+1} |^2 v_{mt}^{x,\ell+1}$
			\State \quad $\bar Z_{nt}^{(1),\ell+1}  =  \sum_{m=1}^{M}  \hat{x}_{ mt}^{ \ell+1} \hat{h}_{ nm}^{b, \ell+1} $
			\State \quad $V_{nt}^{(1),\ell+1} = \bar V_{nt}^{(1),\ell+1}  +  \sum_{m=1}^{M}  v_{mt}^{x,\ell+1}  v_{nm }^{b,\ell+1}$
			\State \quad $ Z_{nt}^{(1),\ell+1} =\bar Z_{nt}^{(1),\ell+1} - \tilde{s}_{nt}^{(1)} \bar V_{nt}^{(1),\ell+1}$
			\State \quad $ v^u_{nt} = \left( \frac{1}{\Sigma^{u}_{nt}} +  \frac{1}{v^0_{u,nt}} \right)^{-1}$
			\State \quad $\hat u_{nt} = v^x_{nt} \left( \frac{R^{u}_{nt}}{\Sigma^{u}_{nt}} +  \frac{\hat u^0}{v^0_u} \right)$
			\State \quad $ v^q_{kn} = \left( \frac{1}{\Sigma^{b}_{kn}} +  \frac{1}{v^0_q} \right)^{-1}$
			\State \quad $\hat q_{kn}  = v^q_{kn} \left( \frac{R^{q}_{kn}}{\Sigma^{q}_{kn}} +  \frac{\hat q_{0} }{v^0_q} \right)$  
			\State  \quad  $\bar V_{kt}^{(2),\ell+1} = \sum_{n=1}^{N}  | \hat{u}_{ nt}^{ \ell+1}  |^2 v_{kn}^{q,\ell+1}
			+ |\hat{q}_{ kn}^{ (2),\ell+1} |^2 v_{nt}^{u,\ell+1}$
			\State  \quad $\bar Z_{kt}^{(2),\ell+1}  =  \sum_{n=1}^{N}  \hat{u}_{ nt}^{ \ell} \hat{q}_{ kn}^{  \ell+1} $
			\State \quad $V_{kt}^{(2),\ell+1} = \bar V_{kt}^{(2),\ell+1}  +  \sum_{n=1}^{N}  v_{nt}^{u,\ell+1}  v_{kn }^{q,\ell+1}$
			\State \quad $ Z_{kt}^{(2),\ell+1} =\bar Z_{kt}^{(2),\ell+1} - \tilde{s}_{kt}^{(2)} \bar V_{kt}^{(2),\ell+1}$
			\State \quad \textbf{Until} $\frac{\left\|\widehat{\mathbf{H}}^{b,\ell} \widehat{\mathbf{X}}^{\ell} -\widehat{\mathbf{H}}^{b,\ell-1} \widehat{\mathbf{X}}^{\ell-1}\right\|_{F}^{2}}{\left\|\widehat{\mathbf{H}}^{b,\ell} \widehat{\mathbf{X}}^{\ell}\right\|_{F}^{2}} \leq \epsilon$ or $\ell=L$ 
			\State \textbf{end for} 
			\Ensure $\widehat{\mathbf{H}}^{b,\ell}$, $\widehat{\mathbf{Q}}^{\ell}$ and $\widehat{\mathbf{X}}^{\ell}$.
		\end{algorithmic} \label{alg:als_forw}
	\end{algorithm}

	\subsubsection{Ambiguities Removal} 
	Given the estimates $\left( \hat{\mathbf{H}}^r, \hat{\mathbf{H}}^b, \hat{\mathbf{X}} \right)$ that the proposed algorithm returns for $\left( \mathbf{H}^r, \mathbf{H}^b, \mathbf{X} \right)$, the following mean squared errors quantify how close the estimates are from the real values:
	\begin{equation} \label{equ:mse}
		\setlength\abovedisplayskip{3pt}%shrink space
		\setlength\belowdisplayskip{3pt} 
		\begin{aligned}
			&\operatorname{MSE}_{r}=\frac{\|\mathbf{H}^r -\hat{\mathbf{H}}^r\|_{F}^{2}}{K N}, \quad  
			\operatorname{MSE}_{b}=\frac{\|\mathbf{H}^b -\hat{\mathbf{H}}^b\|_{F}^{2}}{N M},  \\ 
			&\operatorname{MSE}_{x}=\frac{\|\mathbf{X}  -\hat{\mathbf{X}} \|_{F}^{2}}{M T}.  
		\end{aligned} 
	\end{equation} 
	Note that  in matrix factorization, there is an inherent ambiguity in recovering the couple $\left( \mathbf{H}^b, \mathbf{X} \right)$ and $\left( \mathbf{Q}, \mathbf{U} \right)$. As a matter of fact, for any invertible unitary matrix $\mathbf{C}_1 \in \mathcal{R}^{M \times M}$ and $\mathbf{C}_2 \in \mathcal{R}^{M \times M}$, the couples $\left( \mathbf{H}^b \mathbf{C}_1, \mathbf{C}_1^{-1} \mathbf{X} \right)$ and $\left( \mathbf{Q}\mathbf{C}_2 , \mathbf{C}_2^{-1} \mathbf{U} \right)$ generates the same values as $\left( \mathbf{H}^b, \mathbf{X} \right)$ and $\left( \mathbf{Q}, \mathbf{U} \right)$. The existence of ambiguities undermines the estimation performance of the proposed algorithms, thus, it is necessary to remove or mitigate the inherent ambiguities. Moreover, due to the existence of three unknown matrices, the ambiguity issue in the proposed schem is much more complex than that in the single layer bilinear case, such as BiG-AMP. 
	
	 Inspired by the nonnegative matrix factorization in \cite{6784090}, we present an empirical parameter setting for the identifiability of the proposed algorithms. The proposed schemes have a high probability to give essentially unique estimates if the zero entries of $\mathbf{H}^r$ has the density (number of nonzero entries over the number of entries) less than $(1-\frac{N}{K})$ \cite{6784090}. Moreover, it should be noted that the sparser $\mathbf{H}^b, \mathbf{H}^r$ and $\mathbf{X}$ they are, the lower Cramér–Rao lower bound on the accuracy of the estimates it is \cite{6784090}.  
 
	In order to eliminate scaling ambiguities, the first $K_p$ rows of matrix $\mathbf{H}^r$ are assumed to be known \cite{6177982}.  In addition, to remove the phase ambiguity, we adopt the method in \cite{8879620}. In particular, the transmitted signal is designed to be a full-rank matrix.  However, the cost is unreasonably high to obtain the knowledge of $K_p$ rows in channel matrix  $\mathbf{H}^r$ in a large system. Thus, a lower cost ambiguities removal method is proposed.
		
    Specifically, the portion in channel $\mathbf{H}^r$ can be obtained using bilinear recovery methods with the aid of pilot part in transmitted signals.  Specifically, 
	\begin{equation}\label{equ:amb_1}
		\setlength\abovedisplayskip{3pt}%shrink space
		\setlength\belowdisplayskip{3pt}
		\mathbf{Y}_p \triangleq\mathbf{H}^{r}_p  \mathbf{\Phi}  \mathbf{H}^{b}  \mathbf{X}_p+\mathbf{W}_p,
	\end{equation}
    where $\mathbf{H}^{r}_p \in \mathbb{C}^{K_p\times N}$ and $\mathbf{X}_p\in \mathbb{C}^{M\times T_p}$ are the portion of channel $\mathbf{H}^r$ and transmitted signal $\mathbf{X}$, respectively.  $\mathbf{Y}_p \in  \mathbb{C}^{K_p\times T_p}$ and $\mathbf{W}_p \in  \mathbb{C}^{K_p\times T_p}$ are the portion in the received signal $\mathbf{Y}$ and noise $\mathbf{W}$, respectively. Based on the pilot part, the portion of channel can be estimated using PARAFAC decomposition method or BiG-AMP method. 
    
    PARAFAC method leverages multiple phase matrices $\mathbf{\Phi}$ to decompose the high-dimensional received signals, and it can recover the involved channels through a linear combination of multiple rank-one tensors (the details are referred to our previous work \cite{9366805}). In addition, BiG-AMP is a generalized bilinear matrix recovery method, it estimates the involved channels simultaneously \cite{6898015}. 
%    Although these two recovery methods still encounter  a scaling ambiguity from the convergence point, the ambiguity can be removed with the normalization of the first column of $\mathbf{H}^r_p$, which requires much less cost than obtaining knowledge of $K_p$ rows in  $\mathbf{H}^r$, as shown in our previous work \cite{9417121}. 
    The obtained portion of $\mathbf{H}^r$ can be further used in joint channel estimation and signal recovery, as shown in Alg.~\ref{alg:als_back} and Alg.~\ref{alg:als_forw}.
 
	\subsection{Complexity Analysis}  
	The computational cost of the proposed BAMP algorithm is mainly dominated by componentwise squares of $\mathbf{Q},\mathbf{U},\mathbf{H}^b$ and $\mathbf{X}$. Specifically, in the first layer, the computation complexity is dominated by the computation of $\mathbf{H}^b$ and $\mathbf{X}$ related componentwise squares in line 14$\sim$17 in Alg.~\ref{alg:als_back} and line 5 in Alg.~\ref{alg:als_forw}, which is $\mathcal{O}{\left(NMT\right)}$; and in the second layer, the complexity mainly arises from the computation of $\mathbf{Q}$ and $\mathbf{U}$ related componentwise squares in line 6$\sim$9 in Alg.~\ref{alg:als_back} and line 13 in Alg.~\ref{alg:als_forw}, which is $\mathcal{O}(KNT)$. Thus, the total computational cost of the proposed algorithm is $\mathcal{O}\left((KNT+NMT)L\right)$ with $L$ being the iteration number.

	\section{Simulation Results}\label{sec:simulation}
	In this section, we present computer simulation results for the performance of the proposed two schemes. We have particularly simulated the NMSE using the metrics $\|\mathbf{H}^{b}-\widehat{\mathbf{H}}^{b}\|^{2}\|\mathbf{H}^{b}\|^{-2}$, $\|\mathbf{H}^{r}-\widehat{\mathbf{H}}^{r}\|^{2}\|\mathbf{H}^{r}\|^{-2}$ and $\|\mathbf{X} -\widehat{\mathbf{X}} \|^{2}\|\mathbf{X} \|^{-2}$. The scaling ambiguity of the proposed algorithms has been removed with the aid of the first $K_p$ rows of the channel matrix $\mathbf{H}^{r}$. All normalized mean square error (NMSE) curves were obtained after averaging over $500$ independent Monte Carlo channel realizations.  {The iteration stop criterion $\epsilon=10^{-4}$. } We compare the proposed methods with the state-of-art method, the BiGAMP+least squares (LS) method. Specifically, this method consists of two stages: the first stage estimates two channels $\mathbf{H}^r$ and $\mathbf{H}^b$ based on the pilot part $\mathbf{X}_p \in \mathbb{C}^{M \times T_p}$ using BiGAMP; then the data part $\mathbf{X}_d \in \mathbb{C}^{M \times (T-T_p)}$ is estimated based on the obtained channels using the LS in the second stage. 
	
\begin{figure} \vspace{-2mm}
		\begin{center}
			\includegraphics[width=0.45\textwidth]{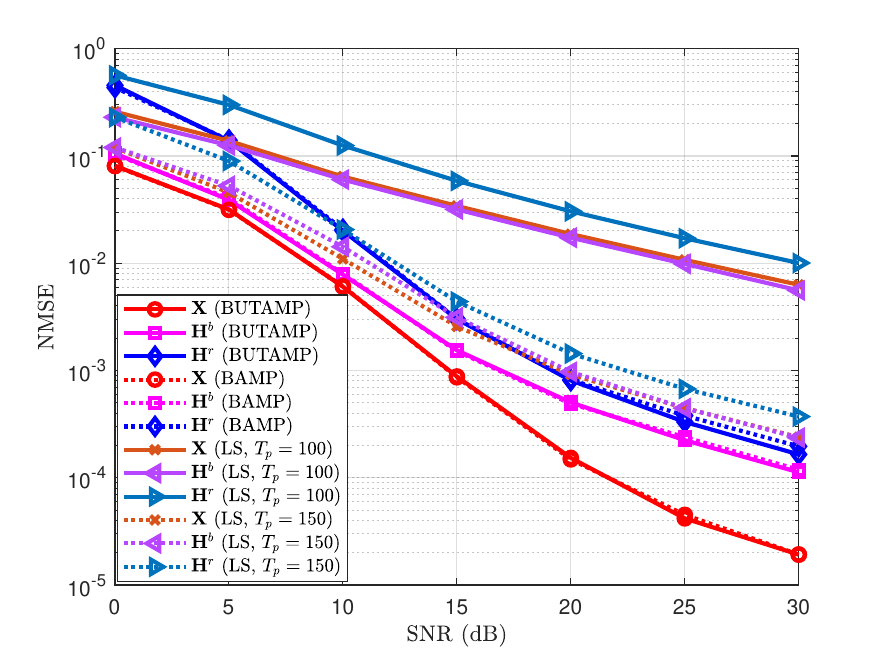}  \vspace{-3mm}
			\caption{NMSE performance comparisons of the two proposed algorithms with baseline method (BiGAMP+LS) versus the SNR in dB for $M=100$, $K=500$, $T=200$ and $N=200$.}
			\label{fig:com_bi_ut} \vspace{-6mm}
		\end{center}
	\end{figure}

	The NMSE performance comparison of the BUTAMP two layers algorithm and BAMP two layers algorithm versus the signal-to-noise ratio (SNR) is given in Fig$.$~\ref{fig:com_bi_ut}. The parameter settings are $M=100$, $K=500$, $T=200$, $N=200$ and $K_p=150$, and the damping factor $\beta$ in BAMP method is set $0.3$. In the proposed algorithms, the pilot length is set $T_p=100$.  It can be observed from the figure that the NMSE performance of the two proposed schemes is similar. In addition, the proposed algorithms consistently show great advantages compared with the benchmark. Specifically, there is  about $18$ dB gap between  the proposed algorithms and the baseline method for the same pilot length $T_p=100$. Even in the unfair setting ($T_p=100$ for the proposed algorithms and  $T_p=150$ for the baseline method), there is about $4$ dB gap between the proposed schemes and the benchmark in estimation of $\mathbf{H}^b$, and the gap in the estimation of $\mathbf{X}$ is even larger.  This behavior	substantiates the favorable performance of our proposed schemes. 

%    The performance evaluation of the BUTAMP two layers algorithm and BAMP  two layers algorithm versus SNR under correlated channels are given in Fig.~\ref{fig:com_cor}. In this figure, $\mathbf{H}^r$ and $\mathbf{H}^b$ are correlated matrix. The parameter settings are $M=100$, $K=500$, $T=200$, $N=200$, $T_p=100$ and $K_p=150$. It can be observed from Fig.~\ref{fig:com_cor} that the proposed two layers algorithms show the similar performance as that of  Fig.~\ref{fig:com_bi_ut} and outperform the baseline method. 
%    \begin{figure} \vspace{-2mm}
%    	\begin{center}
%    		\includegraphics[width=0.7\textwidth]{com_cor.pdf}  \vspace{-3mm}
%    		\caption{NMSE performance comparisons of the BUTAMP algorithm and BAMP algorithm versus the SNR in dB for $M=100$, $K=500$, $T=200$, $N=200$ under correlated Rayleigh fading channels.}
%    		\label{fig:com_cor} \vspace{-6mm}
%    	\end{center}
%    \end{figure}

\begin{figure} \vspace{-2mm}
    	\begin{center}
    		\includegraphics[width=0.45\textwidth]{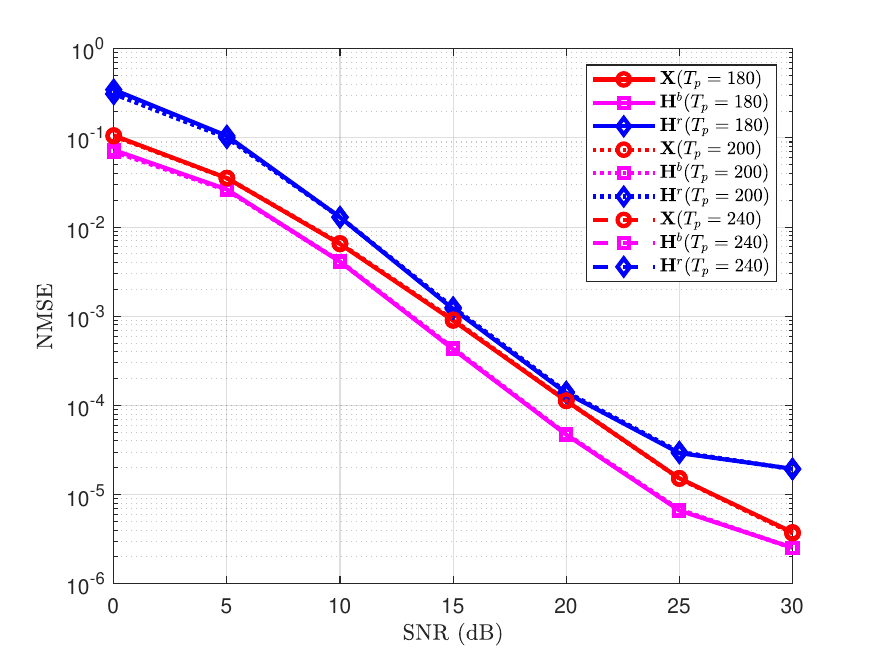}  \vspace{-3mm}
    		\caption{NMSE performance comparisons of the BAMP two layers algorithm versus the SNR in dB for $M=100$, $K=500$, $N=100, T=600$ and various values of $T_p$.}
    		\label{fig:bigamp_com_tp} \vspace{-6mm}
    	\end{center}
    \end{figure}

\begin{figure} \vspace{-2mm}
    	\begin{center}
    		\includegraphics[width=0.45\textwidth]{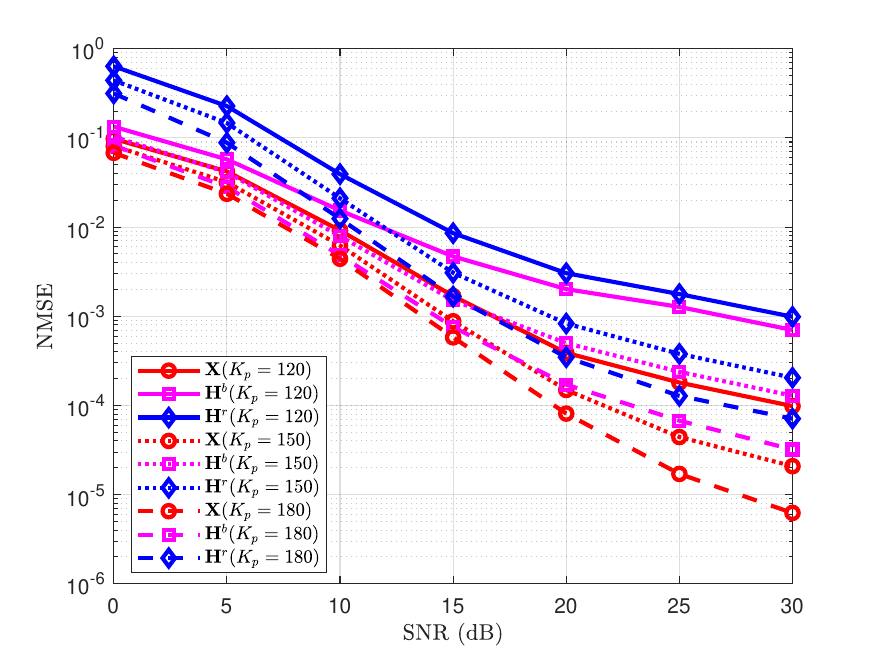}  \vspace{-3mm}
    		\caption{NMSE performance comparisons of the BAMP two layers algorithm versus the SNR in dB for $M=100$, $K=500$, $N=200, T=200$ and various values of $K_p$.}
    		\label{fig:bigamp_com_kp} \vspace{-6mm}
    	\end{center}
    \end{figure}

    We evaluate the influence of pilot part in the Fig$.$~\ref{fig:bigamp_com_tp} and Fig$.$~\ref{fig:bigamp_com_kp}. The minimum number of pilots $T_p$ in $\mathbf{X}$ required for the proposed algrothms is evaluated in Fig$.$~\ref{fig:bigamp_com_tp} with  $M=100$, $K=500$, $N=100, T=600$, and $T_p=180, 200, 240$. It can be observed from the figure that the ratio $T_p/T=0.3$ can achieve the similar performance with the case of higher ratio, which means the ratio of pilots in $\mathbf{X}$ with $0.3$ that is enough to achieve the best performance among all cases. In addition, we evaluate the impact of $K_p$ in $\mathbf{H}^r$ in Fig$.$~\ref{fig:bigamp_com_kp} with $M=100$, $K=500$, $N=100, T=600$, and $K_p=120, 150, 180$. As shown in figure, the larger $K_p$ increases the whole performance. Taking the estimation of $\mathbf{X}$ as an example, the gap between the case with $K_p=120$ and that with $K_p=150$ is about $9$ dB, and the gap between the case with $K_p=150$ and that with $K_p=180$ reduces to $1.5$ dB. The trend of estimations of $\mathbf{H}^r$ and $\mathbf{H}^r$ is similar, and the performance improvement is even larger than the estimation of $\mathbf{X}$, which substantiates that larger $K_p$ could bring more benefit in the proposed algorithm.  

\begin{figure}[h] \vspace{-2mm}
		\begin{center}
			\includegraphics[width=0.45\textwidth]{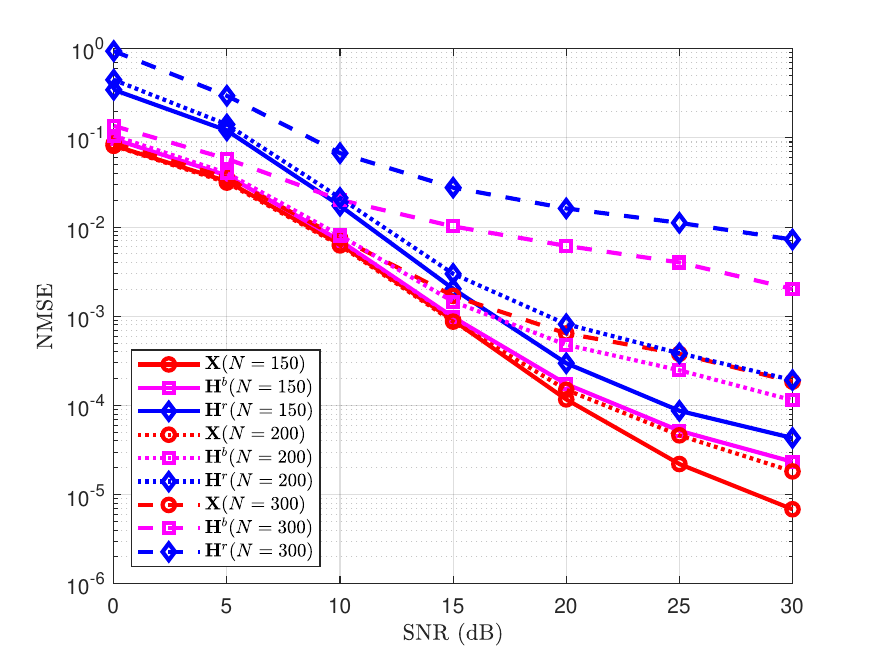}  \vspace{-3mm}
			\caption{NMSE performance comparisons of the BAMP two layers algorithm versus the SNR in dB for $M=100$, $K=500$, $T=200$ and various values of RIS elements $N$.}
			\label{fig:bigamp_com_n} \vspace{-6mm}
		\end{center}
	\end{figure}

	The performance evaluation of the proposed BAMP two layers algorithm versus the SNR with various values of RIS elements $N=150, 200$ and $300$ is given in Fig$.$~\ref{fig:bigamp_com_n}. The parameter settings are $M=100$, $K=500$, $T=200$, $T_p=100$ and $K_p=150$, and the damping factor $\beta$ is set $0.3$. It is evident that there exists an increasing performance loss when $N$ increases, and the gap becomes larger with the increase of SNR, e.g., the gap between the NMSE of $\mathbf{X}$ with $N=150$ and that with $N=200$ is $5$ dB, which is smaller than the gap between cases with $N=150$ and that with $N=300$. In those cases, the number of unknown variables for estimation increases, which results in performance loss.  

\begin{figure} \vspace{-2mm}
    	\begin{center}
    		\includegraphics[width=0.45\textwidth]{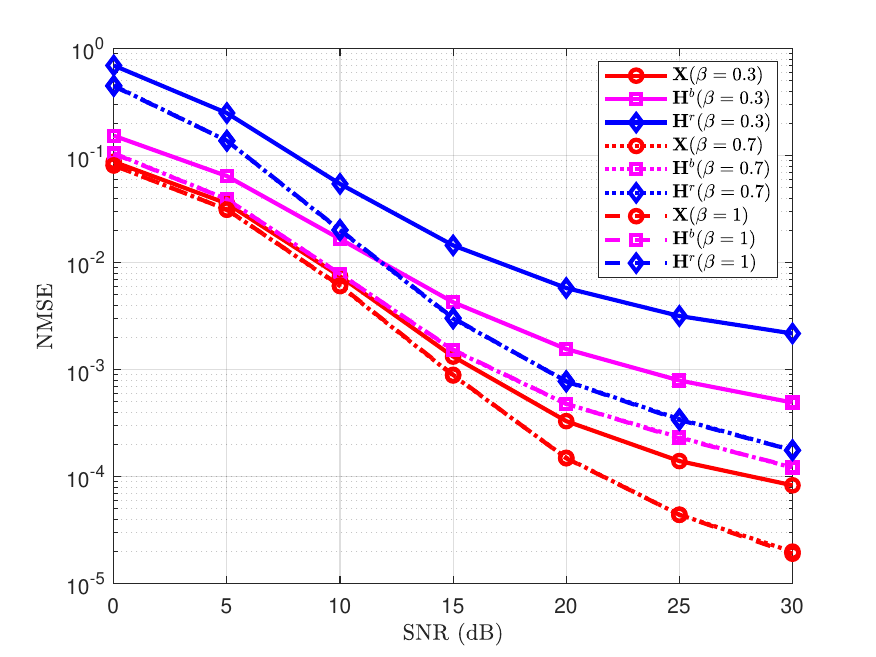}  \vspace{-3mm}
    		\caption{NMSE performance comparisons of the BUTAMP two layers algorithm versus the SNR in dB for $M=100$, $K=500$, $T=200$, $N=200$ and various values of damping factor.}
    		\label{fig:damping_utamp} \vspace{-6mm}
    	\end{center}
    \end{figure}

\begin{figure} \vspace{-2mm}
	 	\begin{center}
	 		\includegraphics[width=0.45\textwidth]{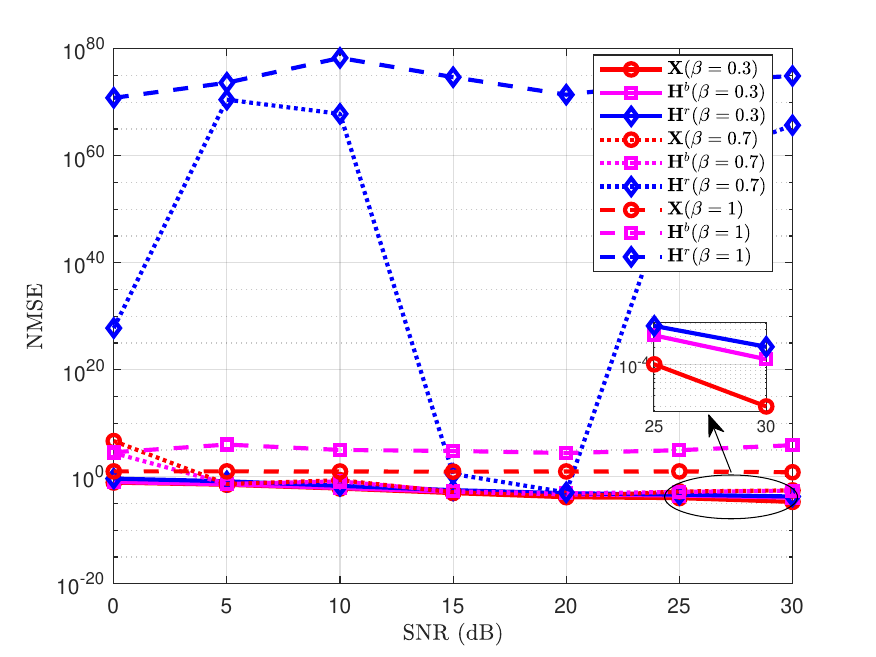}  \vspace{-3mm}
	 		\caption{NMSE performance comparisons of the BAMP two layers algorithm versus the SNR in dB for $M=100$, $K=500$, $T=200$, $N=200$ and various values of damping factor.}
	 		\label{fig:damping_bigamp} \vspace{-6mm}
	 	\end{center}
	 \end{figure}

    The performance evaluation of the BUTAMP two layers algorithm and BAMP two layers algorithm versus SNR with different damping factors are given in Fig.~\ref{fig:damping_utamp} and Fig.~\ref{fig:damping_bigamp}, respectively. The parameter settings are $M=100$, $K=500$, $T=200$, $N=200$, $T_p=100$ and $K_p=150$. It can be observed from Fig.~\ref{fig:damping_utamp} that the BUTAMP two layers algorithm converges with different damping factors, however, the BAMP two layers algorithm diverges easily with inappropriate damping factors. Specifically, the BUTAMP two layers algorithm shows the similar NMSE performance for damping factors $\beta=0.7$ and  $\beta=1$, and all cases converge with the increase of SNR, e.g., the gap of $\mathbf{X}$ estimates between $\beta=0.3$ and $\beta=1$ is about $7$ dB. Different from the Fig.~\ref{fig:damping_utamp}, the NMSE performance of the BAMP two layers algorithm  drastically decreases with changes of damping factor in Fig.~\ref{fig:damping_bigamp}. Specifically, the BAMP two layers algorithm diverges with $\beta=0.7$ and $\beta=1$. By comparing Fig.~\ref{fig:damping_utamp} and Fig.~\ref{fig:damping_bigamp}, we can see that the BUTAMP algorithm is more robust to damping factors. 
	 
\begin{figure}[h] \vspace{-2mm}
	 	\begin{center}
	 		\includegraphics[width=0.45\textwidth]{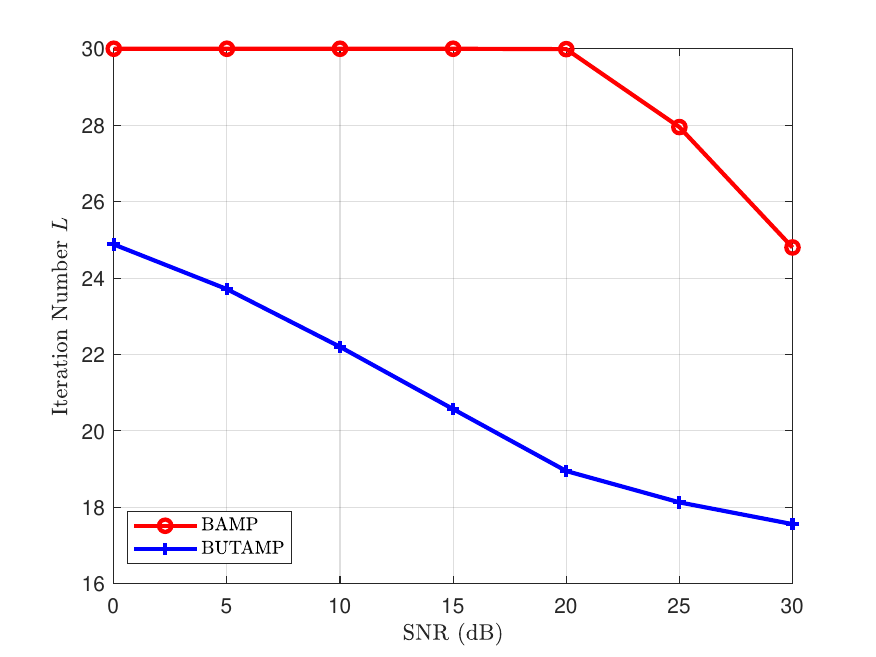}  \vspace{-3mm}
	 		\caption{ Number of iterations versus the SNR in dB for $M=100$, $K=500$, $N=200$, $T=200$ for the proposed BAMP and BUTAMP two layers algorithms. }
	 		\label{fig:com_iteration}\vspace{-6mm}
	 	\end{center}
	 \end{figure}

	 {The convergence of two proposed algorithms is presented in Fig.~\ref{fig:com_iteration}. The parameter setting is $M=100$, $K=500$, $N=200$ and $T=200$. It can be observed from the figure that both algorithms have less number of iterations with the increase of SNR, and BAMP algorithm requires more iterations to converge compared with BUTAMP algorithm, specifically, BAMP algorithm requires $28$ iterations while BUTAMP only requires $18$ iterations at SNR$=20$ dB, which substantiates the favorable convergence performance of BUTAMP algorithm.}
 
	\section{Conclusion} \label{sec:conclusion}
	In this paper, we proposed two message passing schemes for joint channel estimation and signal recovery in RIS-assisted wireless communication systems, which capitalizes on the factor graph and approximate message passing algorithms. All involved channels are estimated and the transmitted signal is recovered through bidirectional two-layer algorithms, BAMP algorithm and BUTAMP algorithm, at the lower computational cost and better convergence performance. Ambiguities removal methods and computational analysis are also presented in this paper. Our simulation results showed that the proposed two schemes have similar performance and BUTAMP algorithm is more robust to damping factors, and two proposed schemes showed superiority over the benchmark scheme. In addition, we observed that the pilot length and the number of RIS elements  exert a significant effect on our proposed algorithm. { Although the proposed algorithms are designed for multiple single-antenna users in uplink, they can be applied in uplink multi-antenna users and downlink single multi-antenna user case. In addition, with the proper design of distributed control networks to exchange information among different users, this work can be extended to downlink multiple multi-antenna users.  }
 
	 \section*{Appendix}
	The message passing between variables and factor nodes in two layers, and the simplification of intermediate variables are derived in this part. 
	\subsubsection{The First Layer (Approximated Factor-to-Variable Messages)}
	We define the approximate posterior distribution of $\mathbf{U}$ as \vspace{-2mm}
	\begin{equation}
		\setlength\abovedisplayskip{3pt}%shrink space
		\setlength\belowdisplayskip{3pt}
		\begin{aligned}
			\xi_{nt}^{u} &=\!p \left(  u_{nt}  \mid   \mathbf{Y} \right) =  x_{m t} \mu_{u_{nt} \!\leftarrow\! h_{nm}^b } (h^b_{nm})  \\
			&+ \!\sum_{m' \neq m}^{M} {\mu_{x_{m' t} \rightarrow f_{u_{nt}}}(x_{m' t})  \mu_{f_{u_{nt}} \leftarrow  h^b_{nm'} } (h^b_{nm'}) }. 
		\end{aligned}
	\end{equation}
	
	Thus, the mean and variance of associate random variable  $\xi_{nt}^u$ are  \vspace{-4mm}
	\begin{equation}
		\setlength\abovedisplayskip{3pt}%shrink space
		\setlength\belowdisplayskip{3pt}
	\begin{aligned}
		\mathbb{E}[\xi_{nt}^u]&=x_{mt} \hat{h}_{nt \leftarrow nm}^b +  \sum_{m' \neq m}^{M} { \hat{h}_{nt \leftarrow nm'}^b  \hat{x}_{m't \rightarrow nt} }\\
		&=x_{mt} \hat{h}_{nt \leftarrow nm}^b + Z_{nt\backslash m},
	\end{aligned}
	\end{equation} \vspace{-2mm}
	\begin{equation}
		\setlength\abovedisplayskip{3pt}%shrink space
		\setlength\belowdisplayskip{3pt}
		\begin{aligned}
			{\rm Var} [\xi_{nt}^u] &= |x_{mt}|^2 v_{nt \leftarrow nm}^{b} + \sum_{m' \neq m}^{M}   |\hat{x}_{m't \rightarrow nt}|^2 v_{nt \leftarrow nm'}^{b} \\
				&+ |\hat{h}_{nt \leftarrow nm'}^b|^2 v_{m't \rightarrow nt}^{x} +  v_{nt \leftarrow nm'}^{b} v_{m't \rightarrow nt}^{x}   \\
			&= |x_{mt}|^2 v_{nt \leftarrow nm}^{b} + V_{nt\backslash m},
		\end{aligned}
	\end{equation} 
	where $\hat{h}_{nt \leftarrow nm}^b$ and $v_{nt \leftarrow nm}^{b}$ are the mean and variance of the message $\mu_{f_{u_{nt}}  \leftarrow h_{nm}^b}$, respectively. $\hat{x}_{m t \rightarrow nt}$ and $v_{m t \rightarrow nt}^{x}$  are the mean and variance of the message $\mu_{x_{m t} \rightarrow f_{u_{nt}} }$.
	
	We define
	\begin{equation}
		\setlength\abovedisplayskip{3pt}%shrink space
		\setlength\belowdisplayskip{3pt}
		\begin{aligned}
			&Z_{nt}^{(1)} = \sum_{m=1}^{M} { \hat{h}_{nt \leftarrow nm }^b  \hat{x}_{mt \rightarrow nt} },  \\
			&V_{nt}^{(1)}  = \sum_{m=1}^{M}  |\hat{x}_{mt \rightarrow nt}|^2 v_{nt \leftarrow nm}^{b} + |\hat{h}_{nt \leftarrow nm}^b|^2 v_{m t \rightarrow nt}^{x}  \\
				&\qquad \qquad+  v_{nt \leftarrow nm}^{b} v_{m t \rightarrow nt}^{x} , 
					\end{aligned}
			\end{equation}
		\begin{equation}\notag
			\setlength\abovedisplayskip{3pt}%shrink space
			\setlength\belowdisplayskip{3pt}
			\begin{aligned}
			&G_{nt}^{(1)}  \left(\mathbb{E}[\xi_{nt}^u],  {\rm Var}[\xi_{nt}^u] \right) \\
			&= \log \int \mathcal{N} \left( u_{nt} \mid \mathbb{E}[\xi_{nt}^u],   {\rm Var}[\xi_{nt}^u] \right)  \mu_{f_{u_{nt}} \leftarrow u_{nt}}  (u_{nt} ) \mathrm{d}{u_{nt}},
		\end{aligned}
	\end{equation}
	the superscript $(i), i=1,2$ denotes the first layer and the second layer, respectively. For notational simplicity, the superscript is dropped in the following derivation. In the large system limits, $\mu_{x_{mt}} (x_{mt})$ is slightly different from $\mu_{x_{mt  \rightarrow nt}} (x_{mt})$. We further use $\hat x_{mt}$ to replace $\hat{x}_{mt \rightarrow nt}$. Besides, the items $ v_{nt \leftarrow nm}^{b} v_{m t \rightarrow nt}^{x} \sim \mathcal{O} (\frac{1}{m})$ and $|\hat{h}_{nt \leftarrow nm}^b|^2 \sim \mathcal{O} (\frac{1}{m})$ are infinitesimal items that can be ignored \cite{zou2020multi}. Thus, we obtain
	\begin{equation}
		\setlength\abovedisplayskip{3pt}%shrink space
		\setlength\belowdisplayskip{3pt}
		\begin{aligned}
			&\log \mu_{x_{mt} \leftarrow f_{u_{nt}}} (x_{mt})  \propto G_{nt} \left(\mathbb{E}[\xi_{nt}^u],  {\rm Var}[\xi_{nt}^u] \right) 
		\end{aligned}
	\end{equation}
\begin{equation}\notag
	\begin{aligned} 
		\setlength\abovedisplayskip{3pt}%shrink space
		\setlength\belowdisplayskip{3pt} 
			&\propto  G_{nt} \left(Z_{nt} \! + \! \hat{h}_{nm}^b  \left(  x_{mt} \!- \!\hat{x}_{mt} \right),  V_{nt} \! + \!   v_{nm}^{b} \left( |x_{mt}|^2 \!- \! |\hat{x}_{mt}|^2\right) \right). 
		\end{aligned}
	\end{equation}
	
	The Taylor series expansion is applied to  $\log \mu_{x_{mt} \leftarrow f_{u_{nt}}} (x_{mt})$
	\begin{equation}\notag
		\setlength\abovedisplayskip{3pt}%shrink space
		\setlength\belowdisplayskip{3pt}
		\begin{aligned}
			&\log \mu_{x_{mt} \leftarrow f_{u_{nt}}} (x_{mt})
		\end{aligned}
	\end{equation}
\begin{equation}
\setlength\abovedisplayskip{3pt}%shrink space
\setlength\belowdisplayskip{3pt}
\begin{aligned}
			&\propto G_{nt} \left( Z_{nt}, V_{nt} \right) +   \hat{h}_{nm}^b  \left(  x_{mt} - \hat{x}_{mt} \right) G'_{nt} \left( Z_{nt}, V_{nt} \right)  \\
			&  +  \frac{ |\hat{h}_{nm}^b|^2   |x_{mt} - \hat{x}_{mt}|^2  }{2}  G''_{nt} \left( Z_{nt}, V_{nt} \right) +  v_{ nm}^{b} \left( |x_{mt}|^2 -  |\hat{x}_{mt}|^2\right) \\
			&\dot{G}_{nt} \left( Z_{nt}, V_{nt} \right)  + \mathcal{O}(\frac{1}{m^{3/2}}) \\
			& \propto x_{mt} \left(  \hat{h}_{nm}^b G'_{nt}\left( Z_{nt}, V_{nt} \right) - |\hat{h}_{nm}^b|^2 \hat{x}_{mt}  G''_{nt}\left( Z_{nt}, V_{nt} \right)
			\right)
			\\ & \quad +  |x_{mt}|^2  \left( \frac{|\hat{h}_{nm}^b|^2}{2} G''_{nt} \left( Z_{nt}, V_{nt} \right) +   v_{nm}^{b} \dot{G}_{nt} \left( Z_{nt}, V_{nt} \right) \right),
		\end{aligned}
	\end{equation}
	where $G'_{nt}$ and $G''_{nt}$ are the first and second partial derivatives of $G_{nt}$ w.r.t. the first argument, and $\dot G_{nt}$ is the first derivative w.r.t. its second argument.

	Thus, we obtain the first derivative of $G_{nt}$
	\begin{equation}
		\setlength\abovedisplayskip{3pt}%shrink space
		\setlength\belowdisplayskip{3pt}
		\begin{aligned}
			&\tilde{s}_{nt} \triangleq G'_{nt}=\frac {\partial G_{nt} \left(Z_{nt},  V_{nt} \right)}  {\partial Z_{nt}}  \\
			&= \frac {\partial\log \int \mathcal{N}  \left( u_{nt} \mid Z_{nt},  V_{nt} \right)  \mu_{f_{u_{nt}} \leftarrow u_{nt}}  (u_{nt} ) \mathrm{d}{u_{nt}}  } {\partial Z_{nt}}  \\
			&=  \frac{  \int \frac{ (u_{nt}-Z_{nt}) } {V_{nt}}  \mathcal{N}  \left( u_{nt} \mid Z_{nt},  V_{nt} \right)  \mu_{f_{u_{nt}} \leftarrow u_{nt}}  (u_{nt} )  \mathrm{d}{u_{nt}}}  {\int \mathcal{N}  \left( u_{nt} \mid Z_{nt},  V_{nt} \right)  \mu_{f_{u_{nt}} \leftarrow u_{nt}}  (u_{nt} ) \mathrm{d}{u_{nt}}} \\
			&= \frac{   \tilde{z}_{nt} - Z_{nt}  }   {V_{nt}},  
		\end{aligned}
	\end{equation}
	where $\frac{   \mathcal{N}  \left( u_{nt}\! \mid\! Z_{nt},  V_{nt} \right)    \mu_{f_{u_{nt}} \leftarrow u_{nt}}  (u_{nt} ) }  {\!\int\! \mathcal{N}\!  \left( u_{nt}\! \mid\! Z_{nt},  V_{nt} \right)  \mu_{f_{u_{nt}} \leftarrow u_{nt}}  (u_{nt} ) \mathrm{d}{u_{nt}}} \! \sim \! \mathcal{N} \!\left( \! u_{nt};  \tilde{z}_{nt} \!,\! \tilde{v}_{nt} \!\right)$.
	
	The second partial derivation $G''_{nt}$ is
	\begin{equation}
		\setlength\abovedisplayskip{3pt}%shrink space
		\setlength\belowdisplayskip{3pt}
		\begin{aligned}
			&v^s_{nt} \triangleq -G''_{nt} = -\frac {\partial^2 G_{nt} \left(Z_{nt},  V_{nt} \right)}  {\partial  Z_{nt}^2}  \\
			&= -\frac {\partial^2\log \int \mathcal{N}  \left( u_{nt} \mid Z_{nt},  V_{nt} \right)  \mu_{f_{u_{nt}} \leftarrow u_{nt}}  (u_{nt} ) \mathrm{d}{u_{nt}}  } {\partial Z_{nt}^2}  
		\end{aligned}
	\end{equation}
\begin{equation}\notag
\begin{aligned} 
\setlength\abovedisplayskip{3pt}%shrink space
\setlength\belowdisplayskip{3pt}
			&= -\partial \! \frac{  \int \frac{  (u_{nt}-Z_{nt}) } {V_{nt}}  \mathcal{N}  \left( u_{nt} \mid Z_{nt},  V_{nt} \right) \!  \mu_{f_{u_{nt}} \!\leftarrow u_{nt}}  (u_{nt} )  \mathrm{d}{u_{nt}}}  {\int \mathcal{N}  \left( u_{nt} \mid Z_{nt},  V_{nt} \right) \! \mu_{f_{u_{nt}} \leftarrow u_{nt}}  (u_{nt} ) \mathrm{d}{u_{nt}}} \! / \!\partial Z_{nt}\\
			& \propto  - \frac{  \tilde{v}_{nt} - V_{nt}  } {V_{nt}^2}.
		\end{aligned}
	\end{equation}
	
	The first partial derivation $\dot G_{nt}$ is given
	\begin{equation}
		\setlength\abovedisplayskip{3pt}%shrink space
		\setlength\belowdisplayskip{3pt}
		\begin{aligned}
			&\dot G_{nt} = \frac {\partial G_{nt} \left(Z_{nt},  V_{nt} \right)}  {\partial V_{nt}} \\
			&=  \frac{  \int  \left[  \frac{ (u_{nt}-Z_{nt})^2 } {2 V_{nt}^2} - \frac {1}{2V_{nt}} \right]  \mathcal{N}  \left( u_{nt} \mid Z_{nt},  V_{nt} \right)  \mu_{f_{u_{nt}} \leftarrow u_{nt}}  (u_{nt} )  \mathrm{d}{u_{nt}}}  {\int \mathcal{N}  \left( u_{nt} \mid Z_{nt},  V_{nt} \right)  \mu_{f_{u_{nt}} \leftarrow u_{nt}}  (u_{nt} ) \mathrm{d}{u_{nt}}} \\ 
			&= \frac {\tilde{z}_{nt}^2 -2 \tilde{z}_{nt} Z_{nt} + Z_{nt}^2 + \tilde{v}_{nt}   - V_{nt}}{ 2 V_{nt}^2}     = \frac{1}{2} \left[ G_{nt}^{\prime \quad 2} + G''_{nt} \right]. 
		\end{aligned}
	\end{equation}
	
	The message can be further simplified as 
	\begin{equation}
		\setlength\abovedisplayskip{3pt}%shrink space
		\setlength\belowdisplayskip{3pt}
		\begin{aligned}
			&\mu_{x_{mt}\! \leftarrow\! f_{u_{nt}}}\! (x_{mt})\! \propto \!\!\!\int\! \!p\! \!\left(\! u_{nt} \!\mid\! x_{mt} h_{nm}^b  \!\!+ \!\!\!\sum_{m' \neq m}^{M}\!\! {x_{m't} h_{nm'}^b }\! \!\!\right) 
		\end{aligned}
	\end{equation}
\begin{equation}\notag
	\setlength\abovedisplayskip{3pt}%shrink space
	\setlength\belowdisplayskip{3pt}
	\begin{aligned}
			&\mu_{f_{u_{nt}} \leftarrow  u_{nt}} \prod_{m'\neq m}^{M} {\mu_{x_{m' t} \rightarrow f_{u_{nt}}}} \prod_{m=1}^{M} {\mu_{f_{u_{nt}} \leftarrow  h^b_{nm} }}  \mathrm{d}{u_{nt}} \mathrm{d}{h^b_{nm}} \\
			&  \sim \mathcal{N} \left(   x_{mt} \mid \frac {\hat{h}_{nm}^b \tilde{s}_{nt} +  \hat{x}_{mt} |\hat{h}_{nm}^b|^2  v^s_{nt}} {|\hat{h}_{nm}^b|^2  v^s_{nt} + v^b_{nm} v^s_{nt} - v^b_{nm} |\tilde{s}_{nt}|^2 } , \right. \\
			&\qquad \qquad \left. \frac{1} {|\hat{h}_{nm}^b|^2  v^s_{nt} + v^b_{nm} v^s_{nt} - v^b_{nm} |\tilde{s}_{nt}|^2 }  \right).
		\end{aligned}
	\end{equation} 

	Similarly, the message from $f_{u_{nt}}$ to $h_{nm}^b$ is given by
	\begin{equation}
		\setlength\abovedisplayskip{3pt}%shrink space
		\setlength\belowdisplayskip{3pt}
		\begin{aligned}
			&\mu_{f_{u_{nt}} \rightarrow h_{nm}^b} (h_{nm}^b) \\
			& =  f_{u_{nt}} \mu_{f_{u_{nt}} \leftarrow  u_{nt}}  \prod_{m'\neq m}^{M} {\mu_{f_{u_{nt}} \leftarrow h_{nm'}^b}}  \prod_{m=1}^{M} \mu_{x_{mt}  \rightarrow {f_{u_{nt}}}}
			\\& \sim \mathcal{N} \left( h_{nm}^b \mid  \frac {\hat{x}_{mt}  \tilde{s}_{nt} +  \hat{h}_{nm}^b |\hat{x}_{mt}^b|^2  v^s_{nt}} {|\hat{x}_{mt}|^2  v^s_{nt} + v^x_{mt} v^s_{nt} - v^x_{mt} |\tilde{s}_{nt}|^2 } , \right. \\ 
			&\qquad \qquad \left. \frac{1} {|\hat{x}_{mt}|^2  v^s_{nt} + v^x_{mt} v^s_{nt} - v^x_{mt} |\tilde{s}_{nt}|^2 }  \right).
		\end{aligned}
	\end{equation}
	
	\subsubsection{The First Layer (Approximated Variable-to-Factor Messages)} 
	The message from $x_{mt}$ to $f_{u_{nt}}$  in $(\ell+1)$-th iteration  is given by \vspace{-0.4cm}
	\begin{equation}
		\setlength\abovedisplayskip{3pt}%shrink space
		\setlength\belowdisplayskip{3pt}
		\mu^{\ell+1}_{x_{mt} \rightarrow f_{u_{nt}}} (x_{mt}) = p(x_{mt}) \prod_{n' \neq n}^{N} {\mu^{\ell}_{x_{mt} \leftarrow f_{u_{n't}}} (x_{mt})},
	\end{equation}  
	where $p(x_{mt})$ is the $(m,t)$-th element of $p(\mathbf{X})$, and  $p(\mathbf{X})\sim \mathcal{N}(\mathbf{x}^0,\mathbf{v}_x^0)$ is a Gaussian mixture, thus we approximate it to be Gaussian with expectation propagation, which is crucial to achieve the low complexity implementation. 
	
	To deal with the discrete valued $x_{mt}$,  the expectation propagation (EP) message $\tilde{p}(x_{mt})$ is used, where the EP message  can be written as \vspace{-0.2cm}
	\begin{equation}
		\setlength\abovedisplayskip{3pt}%shrink space
		\setlength\belowdisplayskip{3pt}
		\begin{aligned}
			\tilde{p}(x_{mt}) \propto \frac{\mu_{x_{mt}}(x_{mt})} {\prod_{n=1}^{N} \mu_{x_{mt}\leftarrow f_{u_{nt}}}}, 
		\end{aligned}
	\end{equation} 
	where $\mu_{x_{mt}}(x_{mt})$ is given in \eqref{equ:poste_x}, and $\prod_{n=1}^{N} \mu_{x_{mt}\leftarrow f_{u_{nt}} (x_{mt})}$ is given in \eqref{equ:RxSigmax}.
	
	Thus, the Gaussian product item in message $\mu _{x_{mt} \rightarrow f_{u_{nt}}} (x_{mt}) $ is  \vspace{-0.2cm}
	\begin{equation}
		\setlength\abovedisplayskip{3pt}%shrink space
		\setlength\belowdisplayskip{3pt}
		\prod_{n'\neq n}^{N} {\mu_{x_{mt} \leftarrow f_{u_{n't}}} (x_{mt})}  \propto \mathcal{N} \left( x_{mt} \mid R^{x}_{mt\backslash n}, \Sigma^{x}_{mt\backslash n} \right),
	\end{equation}
	where  \vspace{-0.4cm}
	\begin{equation}  
		\setlength\abovedisplayskip{3pt}%shrink space
		\setlength\belowdisplayskip{3pt}
		\begin{aligned} 
			&\Sigma^{x}_{mt\backslash n} = \!\left(\! { \!\sum_{n' \neq n}^{N}\! \! |\hat{h}_{n'm}^b|^2  v^s_{n't}\! +\! v^b_{n'm} v^s_{n't}\! - \!v^b_{n'm} \!|\tilde{s}_{n't}|^2 } \!\right) ^{-1}\!\!, \\
			&R^{x}_{mt\backslash n} = \Sigma^{x}_{mt\backslash n}  \left( \sum_{n' \neq n}^{N} {\hat{h}_{n'm}^b \tilde{s}_{n't} +  \hat{x}_{mt} |\hat{h}_{n'm}^b|^2  v^s_{n't} } \right) \\
			&=\hat{x}_{mt} \left( 1+ \Sigma^{x}_{mt\backslash n} \sum_{n' \neq n}^{N} {v^b_{n't} |\tilde{s}_{n't}|^2 - v^b_{n'm} v^s_{n't}}  \right) \\
			&+ \Sigma^{x}_{mt\backslash n}  \sum_{n' \neq n}^{N} {\hat{h}_{n'm}^b \tilde{s}_{n't}}.
		\end{aligned}
	\end{equation}
	
	Thus,
	
	\begin{equation}
		\setlength\abovedisplayskip{3pt}%shrink space
		\setlength\belowdisplayskip{3pt}
	\begin{aligned}
		\mu^{\ell+1}_{x_{mt} \rightarrow f_{u_{nt}}} (x_{mt})& \sim  p(x_{mt}) \mathcal{N} \left( x_{mt} \mid R^{x}_{mt\backslash n}, \Sigma^{x}_{mt\backslash n} \right) \\
		&\sim \mathcal{N} \left( x_{mt \rightarrow  nt} \mid \hat x_{mt \rightarrow  nt}, v^x_{mt \rightarrow  nt}  \right),
	\end{aligned}
	\end{equation}
	where 
	\begin{equation}  
		\setlength\abovedisplayskip{3pt}%shrink space
		\setlength\belowdisplayskip{3pt}
		\begin{aligned}
			\hat x_{mt \rightarrow  nt} &= \frac{1}{C} \int x_{mt} p(x_{mt}) \mathcal{N} \left( x_{mt} \mid R^{x}_{mt\backslash n}, \Sigma^{x}_{mt\backslash n} \right) {\rm d}{x_{mt}} \\
			&= g_{mt} \left( R^{x}_{mt\backslash n}, \Sigma^{x}_{mt\backslash n} \right), \\
			v^x_{mt \rightarrow  nt} &= \frac{1}{C} \int x_{mt}^2 p(x_{mt}) \mathcal{N} \left( x_{mt} \mid R^{x}_{mt\backslash n}, \Sigma^{x}_{mt\backslash n} \right) {\rm d}{x_{mt}} \\
			&-  \hat x_{mt \rightarrow  nt}^2  =   \Sigma^{x}_{mt\backslash n}  g_{mt}^{\prime} \left( R^{x}_{mt\backslash n}, \Sigma^{x}_{mt\backslash n} \right), 
		\end{aligned}
	\end{equation}
	and $C$ is the normalization constant.

	The belief of $x_{mt}$ can be expressed as 
	\begin{equation} \label{equ:poste_x}
		\setlength\abovedisplayskip{3pt}%shrink space
		\setlength\belowdisplayskip{3pt}
	\begin{aligned}
		&\mu_{x_{mt}}  (x_{mt}) = \tilde{p}(x_{mt}) \prod_{n=1}^{N} \mu_{ x_{mt} \leftarrow f_{u_{nt}}  }  (x_{mt}) \\
		&\propto \tilde{p}(x_{mt}) \mathcal{N} \left( x_{mt} \mid R^{x}_{mt}, \Sigma^{x}_{mt} \right) \sim \mathcal{N} \left( x_{mt} \mid \hat x_{mt}, v^x_{mt}  \right),
	\end{aligned}
	\end{equation}
	where 
	\begin{equation} \label{equ:RxSigmax}
		\setlength\abovedisplayskip{3pt}%shrink space
		\setlength\belowdisplayskip{3pt}
		\begin{aligned}
			&\Sigma^{x}_{mt} = \left( { \sum_{n=1}^{N}  |\hat{h}_{n m}^b|^2  v^s_{nt} + v^b_{nm} v^s_{nt} - v^b_{nm} |\tilde{s}_{nt}|^2 } \right) ^{-1}, \\
			&R^{x}_{mt} = \Sigma^{x}_{mt}  \left( \sum_{n=1}^{N} {\hat{h}_{nm}^b \tilde{s}_{nt} +  \hat{x}_{mt} |\hat{h}_{nm}^b|^2  v^s_{nt} } \right)   \\
			%			&=\hat{x}_{mt} \left( 1+ \Sigma^{x}_{mt} \sum_{n=1}^{N} {v^b_{nt} |\tilde{s}_{nt}|^2 - v^b_{nm} v^s_{nt}}  \right) + \Sigma^{x}_{mt}  \sum_{n=1}^{N} {\hat{h}_{nm}^b \tilde{s}_{nt}} \\
			&\quad =\hat{x}_{mt} \left( 1- \Sigma^{x}_{mt} \sum_{n=1}^{N} {  v^b_{nm} v^s_{nt}}  \right) + \Sigma^{x}_{mt}  \sum_{n=1}^{N} {\hat{h}_{nm}^b \tilde{s}_{nt}}, 
		\end{aligned}
	\end{equation}
	and 
	\begin{equation}
		\setlength\abovedisplayskip{3pt}%shrink space
		\setlength\belowdisplayskip{3pt}
		\begin{aligned}
			\hat x_{mt} &= \frac{1}{C} \int x_{mt} p(x_{mt}) \mathcal{N} \left( x_{mt} \mid R^{x}_{mt}, \Sigma^{x}_{mt} \right) {\rm d}{x_{mt}} \\
			&= g_{mt} \left( R^{x}_{mt}, \Sigma^{x}_{mt} \right), \\
			v^x_{mt} &= \frac{1}{C} \int x_{mt}^2 p(x_{mt}) \mathcal{N} \left( x_{mt} \mid R^{x}_{mt}, \Sigma^{x}_{mt} \right) {\rm d}{x_{mt}} \\
			& \qquad -  \hat x_{mt}^2  =   \Sigma^{x}_{mt}  g_{mt}^{\prime} \left( R^{x}_{mt}, \Sigma^{x}_{mt} \right), 
		\end{aligned}
	\end{equation}
	where $g_{mt}^{\prime} \left( R^{x}_{mt}, \Sigma^{x}_{mt} \right)$ is the partial derivation w.r.t. the first argument.
	
	Similarly, the message from $h_{nm}^b$ to $f_{u_{nt}}$  in the $(\ell+1)$-th iteration is given by
	\begin{equation}
		\setlength\abovedisplayskip{3pt}%shrink space
		\setlength\belowdisplayskip{3pt}
		\mu^{\ell+1}_{f_{u_{nt}} \leftarrow h_{nm}^b} ( h_{nm}^b) = p( h_{nm}^b) \prod_{t' \neq t}^{T} {\mu^{\ell}_{ f_{u_{nt'}} \rightarrow h_{nm}^b} } (h_{nm}^b).
	\end{equation}
	
	Thus, the Gaussian product item in message $\mu^{\ell+1}_{f_{u_{nt}} \leftarrow h_{nm}^b} ( h_{nm}^b)$ is 
	\begin{equation}
		\setlength\abovedisplayskip{3pt}%shrink space
		\setlength\belowdisplayskip{3pt}
		\prod_{t' \neq t}^{T} {\mu^{\ell}_{ f_{u_{nt'}} \rightarrow h_{nm}^b} } (h_{nm}^b) \propto \mathcal{N} \left( h_{nm}^b \mid R^{b}_{nm\backslash t}, \Sigma^{b}_{nm\backslash t}  \right),
	\end{equation}
	where 
	\begin{equation}
		\setlength\abovedisplayskip{3pt}%shrink space
		\setlength\belowdisplayskip{3pt}
		\begin{aligned}
			\Sigma^{b}_{nm\backslash t} & \!=\! \left(\! { \sum_{t' \neq t}^{T}  |\hat{x}_{mt'}|^2  v^s_{nt'}\! +\! v^x_{mt'} v^s_{nt'} \!-\! v^x_{mt'} |\tilde{s}_{nt'}|^2  } \!\right) ^{-1}\!,\\
			R^{b}_{nm\backslash t} &\!=\! \Sigma^{b}_{nm\backslash t}  \left( \sum_{t' \neq t}^{T} {\hat{x}_{mt'}  \tilde{s}_{nt'} +  \hat{h}_{nm}^b |\hat{x}_{mt'}^b|^2  v^s_{nt'} } \right).  
		\end{aligned}
	\end{equation}
	
	Thus, 
	
	\begin{equation}
		\setlength\abovedisplayskip{3pt}%shrink space
		\setlength\belowdisplayskip{3pt}
	\begin{aligned}
		\mu^{\ell+1}_{f_{u_{nt}} \leftarrow h_{nm}^b} ( h_{nm}^b)&\! \sim \! p( h_{nm}^b)\! \mathcal{N} \!\left(\! h_{nm}^b \mid R^{b}_{nm\backslash t}, \Sigma^{b}_{nm\backslash t} \! \right) \\
		&\!\sim\! \mathcal{N} \left( h_{nm}^b \mid  \hat{h}_{nt \leftarrow nm}^b, v^b_{nt \leftarrow nm}\right),
	\end{aligned}
	\end{equation}
	where 
	\begin{equation}
		\setlength\abovedisplayskip{3pt}%shrink space
		\setlength\belowdisplayskip{3pt}
		\begin{aligned}
			&\hat{h}_{nt \leftarrow nm}^b = \frac{1}{C} \int h_{nm}^b p(h_{nm}^b) \mathcal{N} \left( h_{nm}^b \mid R^{b}_{nm\backslash t}, \Sigma^{b}_{nm\backslash t} \right) {\rm d}{h_{nm}^b} \\
			&= g_{nm} \left( R^{b}_{nm\backslash t}, \Sigma^{b}_{nm\backslash t} \right), \\
			&v^b_{nt \leftarrow nm} = \frac{1}{C} \int h_{nm}^{b \ 2} p( h_{nm}^b) \mathcal{N} \left( h_{nm}^b \mid R^{b}_{nm\backslash t}, \Sigma^{b}_{nm\backslash t} \right) {\rm d}{ h_{nm}^b} \\
			&-  \hat{h}_{nt \leftarrow nm}^{b \ 2}  =   \Sigma^{b}_{nm\backslash t}  g_{nm}^{\prime} \left( R^{b}_{nm\backslash t}, \Sigma^{b}_{nm\backslash t} \right).
		\end{aligned}
	\end{equation}
	
	The belief of $h_{nm}^b$  
	\begin{equation} \label{equ:poste_b}
		\setlength\abovedisplayskip{3pt}%shrink space
		\setlength\belowdisplayskip{3pt}
		\begin{aligned}
			&\mu_{h_{nm}^b}  (h_{nm}^b) = p(h_{nm}^b) \prod_{t=1}^{T} \mu_{ f_{u_{nt}}  \rightarrow h_{nm}^b }  (h_{nm}^b) \\
			&\sim \!p( h_{nm}^b) \!  \mathcal{N} \!\left(\! h_{nm}^b\! \mid \!R^{b}_{nm}, \Sigma^{b}_{nm}  \!\right) \!\sim\! \mathcal{N} \!\left(\! h_{nm}^b \!\mid\! \hat{h}_{nm}^b, v^b_{nm}\! \right),
		\end{aligned}
	\end{equation}
	where 
	\begin{equation}
		\setlength\abovedisplayskip{3pt}%shrink space
		\setlength\belowdisplayskip{3pt}
		\begin{aligned}
			\Sigma^{b}_{nm } & = \left( { \sum_{t=1}^{T}  |\hat{x}_{mt }|^2  v^s_{nt } + v^x_{mt } v^s_{nt } - v^x_{mt } |\tilde{s}_{nt }|^2  } \right) ^{-1},\\
			R^{b}_{nm } &= \Sigma^{b}_{nm }  \left( \sum_{t=1}^{T} {\hat{x}_{mt }  \tilde{s}_{nt } +  \hat{h}_{nm}^b |\hat{x}_{mt }^b|^2  v^s_{nt } } \right)  \\
			&=\hat{b}_{nm} \left( 1+ \Sigma^{b}_{nm} \sum_{t=1}^{T} {v^x_{mt} |\tilde{s}_{nt}|^2 - v^x_{mt} v^s_{nt}}  \right) \\
			&\qquad + \Sigma^{b}_{nm}  \sum_{t=1}^{T} {\hat{x}_{mt} \tilde{s}_{nt}},  
		\end{aligned}
	\end{equation}
	and 
	\begin{equation}
		\setlength\abovedisplayskip{3pt}%shrink space
		\setlength\belowdisplayskip{3pt}
		\begin{aligned}
			\hat{h}_{ nm}^b &= \frac{1}{C} \int h_{nm}^b p(h_{nm}^b) \mathcal{N} \left( h_{nm}^b \mid R^{b}_{nm }, \Sigma^{b}_{nm } \right) {\rm d}{h_{nm}^b} \\
			&= g_{nm} \left( R^{b}_{nm }, \Sigma^{b}_{nm } \right), \\
			v^b_{ nm} &= \frac{1}{C} \int h_{nm}^{b \ 2} p( h_{nm}^b) \mathcal{N} \left( h_{nm}^b \mid R^{b}_{nm }, \Sigma^{b}_{nm } \right) {\rm d}{ h_{nm}^b} -  \hat{h}_{ nm}^{b \ 2}  \\
			&=   \Sigma^{b}_{nm }  g_{nm}^{\prime} \left( R^{b}_{nm }, \Sigma^{b}_{nm } \right), 
		\end{aligned}
	\end{equation}
	where $g_{nm}^{\prime} \left( R^{b}_{nm}, \Sigma^{b}_{nm} \right)$ is the partial derivation w.r.t. first argument.
	
	\subsubsection{The Second Layer (Approximated Factor-to-Variable Messages)} 
	Similarly, we define the approximate posterior distribution
	\begin{equation}
		\setlength\abovedisplayskip{3pt}%shrink space
		\setlength\belowdisplayskip{3pt}
		\begin{aligned}
			\xi_{kt}^{a}  &=  p \left( a_{kt} \mid \mathbf{Y} \right) = u_{nt} \mu_{f_{a_{kt} \leftarrow q_{kn}}} (q_{kn}) + \\
			&\sum_{n' \neq n}^{N} {\mu_{u_{n't} \rightarrow f_{a_{kt}}} (u_{n't})  \mu_{f_{a_{kt} \leftarrow q_{kn'}}} (q_{kn'}) }.
		\end{aligned}
	\end{equation}
	
	The message from $f_{a_{kt}}$ to $u_{nt}$  can be simplified as
	\begin{equation}
		\setlength\abovedisplayskip{3pt}%shrink space
		\setlength\belowdisplayskip{3pt}
		\begin{aligned}
			&\mu_{u_{nt} \leftarrow f_{a_{kt}} } (u_{nt} )\! = \!\int p\!\left(\!a_{kt} \mid u_{nt} q_{kn} \!+\! \sum_{n' \neq n}^{N} \!{u_{n't} q_{kn'}}\! \right)\!\\
			& \mu_{f_{a_{kt}} \leftarrow a_{kt}  } \prod_{n' \neq n}^{N} {\mu_{ u_{n't} \rightarrow  f_{a_{kt}}} }   \prod_{n=1}^{N} {\mu_{f_{a_{kt}}  \leftarrow  q_{kn}} } \mathrm{d} {a_{kt}} \mathrm{d}{q_{kn}}  \\
			&= \int p\left(a_{kt} \mid \xi_{nt}^{a} \right) \mu_{f_{a_{kt}} \leftarrow a_{kt}} d{a_{kt}}.
		\end{aligned}
	\end{equation}
	
	The mean and variance of the related random variable $\xi_{kt}^a$ can be computed as
	\begin{equation}
		\setlength\abovedisplayskip{3pt}%shrink space
		\setlength\belowdisplayskip{3pt}
		\begin{aligned}
			\mathbb{E} \left[ \xi_{kt}^a \right] &= u_{nt} \hat q_{ kt \leftarrow  kn}+ \sum_{n' \neq n}^{N} {\hat u_{n't \rightarrow kt} \hat q_{kt \leftarrow kn'}} \\
			&=u_{nt} \hat q_{ kt \leftarrow  kn} + Z_{kt \backslash n},
		\end{aligned}	
	\end{equation}
	\begin{equation}
		\setlength\abovedisplayskip{3pt}%shrink space
		\setlength\belowdisplayskip{3pt}
		\begin{aligned}
			{\rm Var }  \left[ \xi_{kt}^a \right] &= |u_{nt}|^2 v_{kt \leftarrow kn}^{q} + \sum_{n' \neq n}^{N}   |\hat u_{n't \rightarrow kt}|^2 v_{kt \leftarrow kn}^{q}\\
				& \qquad + |\hat q_{ kt \leftarrow  kn'}|^2 v_{n't \rightarrow kt}^{u} +  v_{kt \leftarrow  kn'}^{q} v_{n't \rightarrow kt}^{u}   \\
			&= |u_{nt}|^2 v_{kt \leftarrow kn}^{q} + V_{kt\backslash n},
		\end{aligned}
	\end{equation}
	where $\hat{q}_{kt \leftarrow  kn}$ and $v_{kt \leftarrow  kn}^{q}$ are the means and the variance of the message $\mu_{f_{a_{kt}} \leftarrow q_{kn}}$, respectively. In addition, $\hat{u}_{n t \rightarrow kt}$ and $v_{n t \rightarrow kt}^{u}$  are the means and the variance of the message $\mu_{u_{n t} \rightarrow f_{a_{kt}} }$, respectively.
	
	We define
	\begin{equation}
		\setlength\abovedisplayskip{3pt}%shrink space
		\setlength\belowdisplayskip{3pt}
		\begin{aligned}
			&Z_{kt}  \!=\! \sum_{n=1}^{N} { \hat u_{n t \rightarrow kt} \hat q_{kt \leftarrow kn } },  \\
			&V_{kt}  \!=\! \sum_{n=1}^{N} \!\!|\hat u_{n t \rightarrow kt}|^2 v_{kt \leftarrow kn}^{q} \!\!\!+\!\! |\hat q_{ kt \leftarrow  kn }|^2 v_{n t \rightarrow kt}^{u}  \!\!\!+ \! v_{kt \leftarrow  kn }^{q} v_{n t \rightarrow kt}^{u}  .  
		\end{aligned}
	\end{equation}

	Similarly, the logarithm of message $\mu_{u_{nt} \leftarrow f_{a_{kt}} } (u_{nt} )$ is given by
	\begin{equation}
		\setlength\abovedisplayskip{3pt}%shrink space
		\setlength\belowdisplayskip{3pt}
		\begin{aligned}
			&\log \mu_{u_{nt} \leftarrow f_{a_{kt}} } (u_{nt})  \propto G_{kt} \left(\mathbb{E}[\xi_{kt}^a],  {\rm Var}[\xi_{kt}^a] \right)  \\ 
			%			& \propto G_{kt} \left(Z_{kt}-  \hat u_{n t \rightarrow kt} \hat q_{kt \leftarrow kn }   +   u_{n t} \hat q_{kt \leftarrow kn } , \right.  
			%			\\
			%			& \qquad \quad \left. V_{kt} -  |\hat u_{n t \rightarrow kt}|^2 v_{kt \leftarrow kn}^{q} - |\hat q_{ kt \leftarrow  kn }|^2 v_{n t \rightarrow kt}^{u} -  v_{kt \leftarrow  kn }^{q} v_{n t \rightarrow kt}^{u}  +   | u_{n t}|^2 v_{kt \leftarrow  kn }^{q}   \right) \\
			&\propto  G_{kt} \left(Z_{kt} +  \hat{q}_{kn}   \left(  u_{nt} - \hat{u}_{nt} \right),  V_{kt}  +    v_{kt \leftarrow kn}^{q} \left( |u_{nt}|^2 -  |\hat{u}_{nt}|^2\right) \right). 
		\end{aligned}
	\end{equation}
	
	The Taylor series expansion is applied to $\log \!\mu_{u_{nt} \leftarrow \!f_{a_{kt}}}\! \!(u_{nt}) $
	\begin{equation}
		\setlength\abovedisplayskip{3pt}%shrink space
		\setlength\belowdisplayskip{3pt}
		\begin{aligned}
			&\log \mu_{u_{nt} \leftarrow f_{a_{kt}}} (u_{nt})  \propto G_{kt} \left( Z_{kt}, V_{kt} \right) +   \hat{q}_{kn}   \left(  u_{nt} - \hat{u}_{nt} \right) \\
			&  G'_{kt} \left( Z_{kt}, V_{kt} \right)  +  \frac{ |\hat{q}_{kn}|^2   | u_{nt} - \hat{u}_{nt}|^2  }{2}  G''_{kt} \left( Z_{kt}, V_{kt} \right)
			\\ &  +  v_{kt \leftarrow kn}^{q} \left( |u_{nt}|^2 -  |\hat{u}_{nt}|^2\right)  \dot{G}_{kt} \left( Z_{kt}, V_{kt} \right)  + \mathcal{O}(\frac{1}{m^{3/2}}),
		\end{aligned}
	\end{equation}
	where $G'_{kt}$ and $G''_{kt}$ are the first and second partial derivatives of $G_{kt}$ w.r.t. the first argument, and $\dot G_{kt}$ is the first derivative w.r.t. its second argument. 
	
	Thus, we obtain the first derivative of $G_{nt}$
	\begin{equation}
		\setlength\abovedisplayskip{3pt}%shrink space
		\setlength\belowdisplayskip{3pt}
		\begin{aligned}
			\tilde{s}_{kt} \triangleq G'_{kt}&=\frac {\partial G_{kt} \left(Z_{kt},  V_{kt} \right)}  {\partial Z_{kt}} 
			&= \frac{   \tilde{z}_{kt} - Z_{kt}  }   {V_{kt}},  
		\end{aligned}
	\end{equation}
	where $\frac{   \mathcal{N}  \left( a_{kt} \mid Z_{kt},  V_{kt} \right)    \mu_{f_{a_{kt}} \leftarrow a_{kt}}  (a_{kt} ) }  {\int \mathcal{N}  \left( a_{kt} \mid Z_{kt},  V_{kt} \right)  \mu_{f_{a_{kt}} \leftarrow a_{kt}}  (a_{kt} ) \mathrm{d}{a_{kt}}}  \sim  \mathcal{N} \left(  a_{kt};  \tilde{z}_{kt} , \tilde{v}_{kt} \right)$.
	
	The second partial derivation $G''_{kt}$ is 
	\begin{equation}
		\setlength\abovedisplayskip{3pt}%shrink space
		\setlength\belowdisplayskip{3pt}
		\begin{aligned}
			{v^s_{nt}}  \triangleq G'' _{kt} &= \frac {\partial^2 G_{kt} \left(Z_{kt},  V_{kt}\right)}  { {\partial  Z_{kt}}^2 }   
			& \propto   \frac{  \tilde{v}_{kt}  - V_{kt}  } {{V_{kt}}^2}.
		\end{aligned}
	\end{equation}
	
	The first partial derivation $\dot G_{kt} $ is
	\begin{equation}
		\setlength\abovedisplayskip{3pt}%shrink space
		\setlength\belowdisplayskip{3pt}
		\begin{aligned}
			\dot G_{kt}  &= \frac {\partial G_{kt}  \left(Z_{kt},  V_{kt} \right)}  {\partial V_{kt}}     = \frac{1}{2} \left[ {G_{kt}'}^{2} + G''_{kt} \right].
		\end{aligned}
	\end{equation}
	
	Thus, the message from $f_{v_{kt}}$ to $u_{nt}$ can be simplified as
	\begin{equation}
		\setlength\abovedisplayskip{3pt}%shrink space
		\setlength\belowdisplayskip{3pt}
		\begin{aligned}
			\mu _{u_{nt} \!\leftarrow \! f_{v_{kt}}} \!(u_{nt}) & \!\sim\! \mathcal{N}\!\left(\!  \! u_{nt}\! \!\mid \!\!\frac {\hat{q}_{kn} \tilde{s}_{kt} \!+ \! \hat{u}_{nt} |\hat{q}_{kn}|^2  v^s_{kt}} {|\hat{q}_{kn}|^2  v^s_{kt} \!\!+ \!\!v^q_{kn} v^s_{kt} \!-\! v^q_{kn} |\tilde{s}_{nt}|^2 } , \right.\\
			&\left. \frac{1} {|\hat{q}_{kn}|^2  v^s_{kt} + v^q_{kn} v^s_{kt} - v^q_{kn} |\tilde{s}_{kt}|^2 }  \right).
		\end{aligned}
	\end{equation}
	
	Similarly, the message from $f_{a_{kt}}$ to $q_{kn}$  is given by  	
	\begin{equation} 
		\setlength\abovedisplayskip{3pt}%shrink space
		\setlength\belowdisplayskip{3pt}
		\begin{aligned}
			&\mu_{f_{a_{kt}} \rightarrow q_{kn}} (q_{kn})  \\
			&=  f_{a_{kt}} \mu_{f_{a_{kt}} \leftarrow  a_{kt}}  \prod_{n'\neq n}^{N} {\mu_{f_{a_{kt}} \leftarrow q_{kn'}}}  \prod_{n=1}^{N} \mu_{u_{nt}  \rightarrow {f_{a_{kt}}}}  \\
			&\sim \mathcal{N} \left(   q_{kn} \mid \frac {\hat{u}_{nt} \tilde{s}_{kt} +  \hat{q}_{kn} |\hat{u}_{nt}|^2  v^s_{kt}} {|\hat{u}_{nt}|^2  v^s_{kt} + v^u_{nt} v^s_{kt} - v^u_{nt} |\tilde{s}_{nt}|^2 } , \right. \\
			&\qquad \left. \frac{1} {|\hat{u}_{nt}|^2  v^s_{kt} + v^u_{nt} v^s_{kt} - v^u_{nt} |\tilde{s}_{kt}|^2 }  \right).
		\end{aligned}
	\end{equation}
	
	\subsubsection{The Second Layer (Approximated Variable-to-Factor Messages)} 
	The message from $u_{nt}$ to $f_{a_{kt}}$ is given by
	\begin{equation}
		\setlength\abovedisplayskip{3pt}%shrink space
		\setlength\belowdisplayskip{3pt}
	\begin{aligned}
		\mu_{u_{nt} \rightarrow f_{a_{kt}}} (u_{nt}) &\sim  \mu_{f_{u_{nt}} \rightarrow u_{nt} } \mathcal{N} \left( u_{nt} \mid R^{u}_{nt\backslash k}, \Sigma^{u}_{nt\backslash k} \right) \\
		&\sim \mathcal{N} \left( u_{nt \rightarrow  kt} \mid \hat u_{nt \rightarrow  kt}, v^u_{nt \rightarrow  kt}  \right),
	\end{aligned}
	\end{equation}
	where
	\begin{equation}
		\setlength\abovedisplayskip{3pt}%shrink space
		\setlength\belowdisplayskip{3pt}
		\begin{aligned}
			\Sigma^{u}_{nt\backslash k} &= \left( { \sum_{t' \neq t}^{T}  |\hat{q}_{kn}|^2  v^s_{kt'} + v^q_{kn} v^s_{kt'} - v^q_{kn} |\tilde{s}_{kt'}|^2 } \right) ^{-1},  \\
			R^{u}_{nt\backslash k} &= \Sigma^{u}_{nt\backslash k}  \left( \sum_{t' \neq t}^{T} {\hat{q}_{kn} \tilde{s}_{kt'} +  \hat{u}_{nt'} |\hat{q}_{kn}|^2  v^s_{kt'} } \right).
		\end{aligned}
	\end{equation}
	
	Thus, the belief of $u_{nt}$ is
	\begin{equation} 
		\setlength\abovedisplayskip{3pt}%shrink space
		\setlength\belowdisplayskip{3pt}
		\begin{aligned}
			&\mu_{u_{nt} } (u_{nt}) \sim  \mu_{f_{u_{nt}} \rightarrow u_{nt} } \mathcal{N} \left( u_{nt} \mid R^{u}_{nt }, \Sigma^{u}_{nt } \right) \\
			&\sim \mathcal{N} \left( u_{nt } \mid \hat u_{nt }, v^u_{nt }  \right).
		\end{aligned}
	\end{equation}
	
	Similarly,
	\begin{equation}
		\setlength\abovedisplayskip{3pt}%shrink space
		\setlength\belowdisplayskip{3pt}
		\begin{aligned}
			\mu _{f_{a_{kt}} \leftarrow q_{kn} } ( q_{kn}) &\sim  p( q_{kn}) \mathcal{N} \left( q_{kn} \mid R^{q}_{kn \backslash t}, \Sigma^{q}_{kn\backslash t}  \right) \\
			&\sim \mathcal{N} \left( q_{kn} \mid  \hat{q}_{kt \leftarrow kn}, v^q_{kt \leftarrow kn}\right),
		\end{aligned}
	\end{equation}
	and the belief of $q_{kn}$  is given by
	\begin{equation} \label{equ:poste_r}
		\setlength\abovedisplayskip{3pt}%shrink space
		\setlength\belowdisplayskip{3pt}
	\begin{aligned}
		&\mu_{q_{kn}}  (q_{kn})\\
		& p(q_{kn}) \prod_{t=1}^{T} \mu_{ f_{a_{kt}}  \rightarrow q_{kn} }  (q_{kn}) \sim p(q_{kn}) \mathcal{N} \left( q_{kn} \mid R^{q}_{kn}, \Sigma^{q}_{kn}  \right) \\
		&\sim \mathcal{N} \left( q_{kn} \mid \hat{q}_{kn}, v^q_{kn} \right).
	\end{aligned}
	\end{equation}
	
	\subsubsection{Simplications of Intermediate Variables}
	We further simplify   
	\begin{equation}
		\setlength\abovedisplayskip{3pt}%shrink space
		\setlength\belowdisplayskip{3pt}
		\begin{aligned}\notag
			Z_{kt}^{\ell} &= \sum_{n=1}^{N} { \hat u_{n t \rightarrow kt}^{\ell} \hat q_{kt \leftarrow kn }^{\ell} }  = \sum_{n=1}^{N}  \left(\hat{u}_{ nt}^{ \ell}- {\hat{q}_{kn}}^{ \ell-1} \tilde{s}_{kt}^{\ell-1} {v_{nt}^{u,\ell}} \right) 
		\end{aligned}
	\end{equation}
\begin{equation}
	\setlength\abovedisplayskip{3pt}%shrink space
	\setlength\belowdisplayskip{3pt}
	\begin{aligned}     
			&\quad \left( \hat{q}_{ kn}^{ \ell}- {\hat{u}_{nt}}^{ \ell-1} \tilde{s}_{kt}^{\ell-1} {v_{kn}^{q,\ell}} \right)      \\
			&\approx \sum_{n=1}^{N}  \hat{u}_{ nt}^{ \ell} \hat{q}_{ kn}^{ \ell} - \tilde{s}_{kt}^{\ell-1} \sum_{n=1}^{N} \left(   |{\hat{q}_{kn}}^{ \ell}|^2  {v_{nt}^{u,\ell}} +  |\hat{u}_{ nt}^{ \ell}|^2  {v_{kn}^{q,\ell}}   \right) \\
			&=\bar Z_{kt}^{\ell} - \tilde{s}_{kt}^{\ell-1} \bar V_{kt}^{\ell}.
		\end{aligned}
	\end{equation}
	
	Moreover, 
	\begin{equation}
		\setlength\abovedisplayskip{3pt}%shrink space
		\setlength\belowdisplayskip{3pt}
		\begin{aligned}
			V_{kt}^{\ell} &= \sum_{n=1}^{N} {|\hat u_{n t \rightarrow kt}^{\ell}|^{2} v_{kt \leftarrow kn}^{q,\ell} + |\hat q_{ kt \leftarrow  kn }^{\ell}|^2 v_{n t \rightarrow kt}^{u,\ell} +  v_{kt \leftarrow  kn }^{q,\ell} v_{n t \rightarrow kt}^{u,\ell}   } \\
			&=  \sum_{n=1}^{N}  | \hat{u}_{ nt}^{ \ell}- {\hat{q}_{kn}}^{ \ell-1} \tilde{s}_{kt}^{\ell-1} {v_{nt}^{u,\ell}} |^2 v_{kn}^{q,\ell}
			+ |\hat{q}_{ kn}^{ \ell}\\
			&\qquad - {\hat{u}_{nt}}^{ \ell-1} \tilde{s}_{kt}^{\ell-1} {v_{kn}^{q,\ell}} |^2 v_{nt}^{u,\ell}
			+ v_{kn }^{q,\ell} v_{n t}^{u,\ell}  \\
			&\approx 	 \bar V_{kt}^{\ell}  +  \sum_{n=1}^{N}  v_{nt}^{u,\ell}  v_{kn }^{q,\ell}.
		\end{aligned}
	\end{equation}
	
	Similarly,
	\begin{equation}
		\setlength\abovedisplayskip{3pt}%shrink space
		\setlength\belowdisplayskip{3pt}
		\begin{aligned}
			Z_{nt}^{\ell} &= \sum_{m=1}^{M} { \hat{h}_{nt \leftarrow nm }^b  \hat{x}_{mt \rightarrow nt} }  \approx \bar Z_{nt}^{\ell} - \tilde{s}_{nt}^{\ell-1} \bar V_{nt}^{\ell}, \\
			V_{nt}^{\ell} &\approx 	 \bar V_{nt}^{\ell}  +  \sum_{m=1}^{M}  v_{mt}^{x,\ell}  v_{nm }^{b,\ell}.
		\end{aligned}
	\end{equation}
	
	We then simplify
	\begin{equation}
		\setlength\abovedisplayskip{3pt}%shrink space
		\setlength\belowdisplayskip{3pt}
		\begin{aligned}
		\Sigma^{x}_{mt}  &= \left( { \sum_{n=1}^{N}  |\hat{h}_{n m}^b|^2  v^s_{nt} + v^b_{nt} v^s_{nt} - v^b_{nt} |\tilde{s}_{nt}|^2 } \right) ^{-1} 	\\
		&  \approx  \left( { \sum_{n=1}^{N}  |\hat{h}_{n m}^b|^2  v^s_{nt}  } \right) ^{-1},
		\end{aligned}
	\end{equation}
	\begin{equation}
		\setlength\abovedisplayskip{3pt}%shrink space
		\setlength\belowdisplayskip{3pt}
		\begin{aligned}
			\Sigma^{u}_{nt}  \approx  \left( { \sum_{k=1}^{K}  |\hat{q}_{kn}|^2  v^s_{kt}  } \right) ^{-1}.
		\end{aligned}
	\end{equation}
	
	\subsubsection{Message Between Different Layers}
	The message from $u_{nt}$ to $f_{u_{nt}}$ in $\ell$-th iteration is given by
	\begin{equation}
		\setlength\abovedisplayskip{3pt}%shrink space
		\setlength\belowdisplayskip{3pt}
		\mu^{\ell}_{f_{u_{nt}} \leftarrow u_{nt} } (u_{nt}) = \prod_{k=1}^{K} {\mu^{\ell}_{u_{nt} \leftarrow f_{a_{kt} } } },
	\end{equation}
	which is the product of large number of Gaussian distributions. By the Gaussian product property, we obtain
	\begin{equation}
		\setlength\abovedisplayskip{3pt}%shrink space
		\setlength\belowdisplayskip{3pt}
		\mu^{\ell}_{f_{u_{nt}} \leftarrow u_{nt} } (u_{nt}) = \mathcal{N} \left(  u_{nt} \mid R_{nt}, \Sigma_{nt} \right),
	\end{equation}
	where
	\begin{equation}
		\setlength\abovedisplayskip{3pt}%shrink space
		\setlength\belowdisplayskip{3pt}
		\begin{aligned}
			\Sigma_{nt} &= \left(  \sum_{k=1}^{K} \frac{1}{\Sigma_{kt}^u} \right) ^{-1},  R_{nt} &= \Sigma_{nt} \left(  \sum_{k=1}^{K} \frac{R_{kt}^u}{\Sigma_{kt}^u} \right),
		\end{aligned}
	\end{equation}
	with $\mu_{u_{nt} \leftarrow f_{a_{kt}} } \left(u_{nt}\right) \sim \mathcal{N} \left(u_{nt} \mid R_{kt}^u, \Sigma_{kt}^u \right)$.
	
	Thus, 
	$\zeta _{nt} = \frac{   \mathcal{N}  \left( u_{nt} \mid Z_{nt},  V_{nt} \right)    \mu_{f_{u_{nt}} \leftarrow u_{nt}}  (u_{nt} ) }  {\int \mathcal{N}  \left( u_{nt} \mid Z_{nt},  V_{nt} \right)  \mu_{f_{u_{nt}} \leftarrow u_{nt}}  (u_{nt} ) \mathrm{d}{u_{nt}}}  \sim  \mathcal{N} \left(  u_{nt};  \tilde{z}_{nt} , \tilde{v}_{nt} \right)$ can be further updated as 
	\begin{equation} \label{equ:poste_u}
		\setlength\abovedisplayskip{3pt}%shrink space
		\setlength\belowdisplayskip{3pt}
		\begin{aligned}
			\zeta _{nt} &= \frac{   \mathcal{N}  \left( u_{nt} \mid Z_{nt},  V_{nt} \right)   \mathcal{N} \left(  u_{nt} \mid R_{nt}, \Sigma_{nt} \right) }  {\int \mathcal{N}  \left( u_{nt} \mid Z_{nt},  V_{nt} \right)  \mathcal{N} \left(  u_{nt} \mid R_{nt}, \Sigma_{nt} \right) \mathrm{d}{u_{nt}}}  \\
			&\sim  \mathcal{N} \left(  u_{nt};  \tilde{z}_{nt} , \tilde{v}_{nt} \right),
		\end{aligned}
	\end{equation}
	where 
	\begin{equation}
		\setlength\abovedisplayskip{3pt}%shrink space
		\setlength\belowdisplayskip{3pt}
		\begin{aligned}
			\tilde{v}_{nt} &=  \left( \frac{1}{V_{nt}} + \frac{1}{\Sigma_{nt}} \right) ^{-1}, 
			\tilde{z}_{nt} &= \tilde{v}_{nt}  \left( \frac{Z_{nt}}{V_{nt}} + \frac{R_{nt}}{\Sigma_{nt}} \right).
		\end{aligned}
	\end{equation}
	
	Similarly, 
	\begin{equation} \label{equ:poste_a}
		\setlength\abovedisplayskip{3pt}%shrink space
		\setlength\belowdisplayskip{3pt}
		\begin{aligned}
			&\zeta _{kt} = \frac{   \mathcal{N}  \left( a_{kt} \mid Z_{kt},  V_{kt} \right)   \mathcal{N} \left(  a_{kt} \mid R_{kt}, \Sigma_{kt} \right) }  {\int \mathcal{N}  \left( a_{kt} \mid Z_{kt},  V_{kt} \right)  \mathcal{N} \left(  a_{kt} \mid R_{kt}, \Sigma_{kt} \right) \mathrm{d}{a_{kt}}}  \\
			&\sim  \mathcal{N} \left(  a_{kt};  \tilde{z}_{kt} , \tilde{v}_{kt} \right),
		\end{aligned}
	\end{equation}
	where 
	\begin{equation} 
		\setlength\abovedisplayskip{3pt}%shrink space
		\setlength\belowdisplayskip{3pt}
		\begin{aligned}
			\tilde{v}_{kt} &=  \left( \frac{1}{V_{kt}} + \frac{1}{\Sigma_{kt}} \right) ^{-1},  
			\tilde{z}_{kt} &= \tilde{v}_{kt}  \left( \frac{Z_{kt}}{V_{kt}} + \frac{R_{kt}}{\Sigma_{kt}} \right).
		\end{aligned}
	\end{equation}

	\bibliographystyle{IEEEbib}
	\bibliography{strings}

\begin{thebibliography}{10}

\bibitem{Akyildiz2018mag}
I.~F. {Akyildiz}, C.~{Han}, and S.~{Nie},
\newblock ``Combating the distance problem in the millimeter wave and terahertz
  frequency bands,''
\newblock {\em IEEE Commun. Mag.}, vol. 56, no. 6, pp. 102--108, June 2018.

\bibitem{hu2018beyond}
S.~Hu, F.~Rusek, and O.~Edfors,
\newblock ``Beyond massive {MIMO}: The potential of positioning with large
  intelligent surfaces,''
\newblock {\em IEEE Trans. Signal Process.}, vol. 66, no. 7, pp. 1761--1774,
  May 2018.

\bibitem{huang2019reconfigurable}
C.~Huang, A.~Zappone, G.~C. Alexandropoulos, M.~Debbah, and C.~Yuen,
\newblock ``Reconfigurable intelligent surfaces for energy efficiency in
  wireless communication,''
\newblock {\em IEEE Trans. Wirel. Commun.}, vol. 18, no. 8, pp. 4157--4170,
  Aug. 2019.

\bibitem{Marco2019}
M.~Di Renzo, M.~Debbah, D.-T. Phan-Huy, A.~Zappone, M.-S. Alouini, C.~Yuen,
  V.~Sciancalepore, G.~C. Alexandropoulos, J.~Hoydis, H.~Gacanin, J.~de~Rosny,
  A.~Bounceur, G.~Lerosey, and M.~Fink,
\newblock ``Smart radio environments empowered by reconfigurable {AI}
  meta-surfaces: An idea whose time has come,''
\newblock {\em EURASIP J. Wireless Commun. Netw.}, vol. 2019, no. 1, pp. 1--20,
  May 2019.

\bibitem{qingqing2019towards}
Q.~Wu and R.~Zhang,
\newblock ``Towards smart and reconfigurable environment: Intelligent
  reflecting surface aided wireless network,''
\newblock {\em IEEE Commun. Mag.}, vol. 58, no. 1, Jan. 2020.

\bibitem{strinati2021wireless}
E.~{Calvanese Strinati et al.},
\newblock ``Wireless environment as a service enabled by reconfigurable
  intelligent surfaces: The {RISE-6G} perspective,''
\newblock {\em Proc. Joint EuCNC \& { 6G} Summit}, Porto, Portugal, 8–11 June
  2021.

\bibitem{9140329}
M.~D.~Renzo, A.~Zappone, M.~Debbah, M.-S. Alouini, C.~Yuen, J.~de~Rosny, and
  S.~Tretyakov,
\newblock ``Smart radio environments empowered by reconfigurable intelligent
  surfaces: How it works, state of research, and the road ahead,''
\newblock {\em IEEE J. Sel. Areas Commun.}, vol. 38, no. 11, pp. 2450--2525,
  Nov. 2020.

\bibitem{9371416}
B.~Yang, X.~Cao, C.~Huang, C.~Yuen, L.~Qian, and M.~Di Renzo,
\newblock ``Intelligent spectrum learning for wireless networks with
  reconfigurable intelligent surfaces,''
\newblock {\em IEEE Trans. Veh. Tech.}, vol. 70, no. 4, pp. 3920--3925, Apr.
  2021.

\bibitem{9206044}
W.~{Tang}, M.~Z. {Chen}, X.~{Chen}, J.~Y. {Dai}, Y.~{Han}, M.~{Di Renzo},
  Y.~{Zeng}, S.~{Jin}, Q.~{Cheng}, and T.~J. {Cui},
\newblock ``Wireless communications with reconfigurable intelligent surface:
  Path loss modeling and experimental measurement,''
\newblock {\em IEEE Trans. Wirel. Commun.}, vol. 20, no. 1, pp. 421--439, Jan.
  2021.

\bibitem{9136592}
C.~Huang, S.~Hu, G.~C. Alexandropoulos, A.~Zappone, C.~Yuen, R.~Zhang, M.~Di
  Renzo, and M.~Debbah,
\newblock ``Holographic {MIMO} surfaces for {6G} wireless networks:
  Opportunities, challenges, and trends,''
\newblock {\em IEEE Wirel. Commun.}, vol. 27, no. 5, pp. 118--125, Oct. 2020.

\bibitem{alexandropoulos2021reconfigurable}
G.~C. {Alexandropoulos}, N.~{Shlezinger}, and P.~del Hougne,
\newblock ``Reconfigurable intelligent surfaces for rich scattering wireless
  communications: Recent experiments, challenges, and opportunities,''
\newblock {\em IEEE Commun. Mag.}, vol. 59, no. 6, pp. 28–34, June 2021.

\bibitem{husha_LIS2}
S.~{Hu}, F.~{Rusek}, and O.~{Edfors},
\newblock ``Beyond massive {MIMO}: The potential of data-transmission with
  large intelligent surfaces,''
\newblock {\em IEEE Trans. Signal Process.}, vol. 66, no. 10, pp. 2746--2758,
  May 2018.

\bibitem{9110869}
C.~Huang, R.~Mo, and C.~Yuen,
\newblock ``Reconfigurable intelligent surface assisted multiuser {MISO}
  systems exploiting deep reinforcement learning,''
\newblock {\em IEEE J. Sel. Areas Commun.}, vol. 38, no. 8, pp. 1839--1850,
  Aug. 2020.

\bibitem{han2019}
Y.~Han, W.~Tang, S.~Jin, C.~Wen, and X.~Ma,
\newblock ``Large intelligent surface-assisted wireless communication
  exploiting statistical {CSI},''
\newblock {\em IEEE Trans. Veh. Technol.}, vol. 68, no. 8, pp. 8238--8242, Aug.
  2019.

\bibitem{8941126}
W.~{Yan}, X.~{Yuan}, and X.~{Kuai},
\newblock ``Passive beamforming and information transfer via large intelligent
  surface,''
\newblock {\em IEEE Wirel. Commun. Lett.}, vol. 9, no. 4, pp. 533--537, Apr.
  2020.

\bibitem{9133094}
T.~{Hou}, Y.~{Liu}, Z.~{Song}, X.~{Sun}, Y.~{Chen}, and L.~{Hanzo},
\newblock ``Reconfigurable intelligent surface aided {NOMA} networks,''
\newblock {\em IEEE J. Sel. Area. Comm.}, vol. 38, no. 11, pp. 2575--2588, Nov.
  2020.

\bibitem{9318531}
Z.~Peng, Z.~Zhang, C.~Pan, Li~Li, and A.~Lee S.,
\newblock ``Multiuser full-duplex two-way communications via intelligent
  reflecting surface,''
\newblock {\em IEEE Trans. Sig. Process.}, vol. 69, pp. 837--851, Jan. 2021.

\bibitem{9263383}
C.~Pradhan, A.~Li, L.~Song, J.~Li, B.~Vucetic, and Y.~Li,
\newblock ``Reconfigurable intelligent surface ({RIS})-enhanced two-way {OFDM}
  communications,''
\newblock {\em IEEE Trans. Veh. Tech.}, vol. 69, no. 12, pp. 16270--16275, Dec.
  2020.

\bibitem{Alkhateeb2019}
A.~Taha, M.~Alrabeiah, and A.~Alkhateeb,
\newblock ``Enabling large intelligent surfaces with compressive sensing and
  deep learning,''
\newblock {\em IEEE Access}, vol. 9, pp. 44304--44321, Mar. 2021.

\bibitem{9410457}
C.~Huang, Z.~Yang, G.~C. Alexandropoulos, K.~Xiong, L.~Wei, C.~Yuen, Z.~Zhang,
  and M.~Debbah,
\newblock ``Multi-hop {RIS}-empowered terahertz communications: A {DRL}-based
  hybrid beamforming design,''
\newblock {\em IEEE J. Sel. Areas Commun.}, vol. 39, no. 6, pp. 1663--1677,
  June 2021.

\bibitem{9366805}
L.~{Wei}, C.~{Huang}, G.~C. {Alexandropoulos}, C.~{Yuen}, Z.~{Zhang}, and
  M.~{Debbah},
\newblock ``Channel estimation for {RIS}-empowered multi-user {MISO} wireless
  communications,''
\newblock {\em IEEE Trans. Commun.}, pp. 1--1, 2021.

\bibitem{9133142}
C.~{You}, B.~{Zheng}, and R.~{Zhang},
\newblock ``Channel estimation and passive beamforming for intelligent
  reflecting surface: Discrete phase shift and progressive refinement,''
\newblock {\em IEEE J. Sel. Areas Commun.}, vol. 38, no. 11, pp. 2604--2620,
  Nov. 2020.

\bibitem{9417121}
L.~Wei, C.~Huang, G.~C. Alexandropoulos, Z.~Yang, C.~Yuen, and Z.~Zhang,
\newblock ``Joint channel estimation and signal recovery in {RIS}-assisted
  multi-user {MISO} communications,''
\newblock in {\em 2021 IEEE WCNC}, 2021, pp. 1--6.

\bibitem{6898015}
J.~T. {Parker}, P.~{Schniter}, and V.~{Cevher},
\newblock ``Bilinear generalized approximate message passing—part i:
  Derivation,''
\newblock {\em IEEE Trans. Signal Process.}, vol. 62, no. 22, pp. 5839--5853,
  Nov. 2014.

\bibitem{8357527}
X.~{Meng} and J.~{Zhu},
\newblock ``A generalized sparse bayesian learning algorithm for 1-bit {DOA}
  estimation,''
\newblock {\em IEEE Commun. Lett.}, vol. 22, no. 7, pp. 1414--1417, July 2018.

\bibitem{7574287}
S.~{Wu}, L.~{Kuang}, Z.~{Ni}, D.~{Huang}, Q.~{Guo}, and J.~{Lu},
\newblock ``Message-passing receiver for joint channel estimation and decoding
  in {3D} massive {MIMO-OFDM} systems,''
\newblock {\em IEEE Trans. Wirel. Commun.}, vol. 15, no. 12, pp. 8122--8138,
  Dec. 2016.

\bibitem{9103622}
Y.~{Zhang}, Z.~{Yuan}, Q.~{Guo}, Z.~{Wang}, J.~{Xi}, and Y.~{Li},
\newblock ``Bayesian receiver design for grant-free {NOMA} with message passing
  based structured signal estimation,''
\newblock {\em IEEE Trans. Veh. Technol.}, vol. 69, no. 8, pp. 8643--8656, Aug.
  2020.

\bibitem{9006927}
W.~{Yuan}, N.~{Wu}, Q.~{Guo}, D.~W.~K. {Ng}, J.~{Yuan}, and L.~{Hanzo},
\newblock ``Iterative joint channel estimation, user activity tracking, and
  data detection for {FTN-NOMA} systems supporting random access,''
\newblock {\em IEEE Trans. Commun.}, vol. 68, no. 5, pp. 2963--2977, May 2020.

\bibitem{zou2020multi}
Q.~Zou, H.~Zhang, and H.~Yang,
\newblock ``Multi-layer bilinear generalized approximate message passing,''
\newblock {\em IEEE Trans. Sig. Process.}, vol. 69, pp. 4529--4543, Jul. 2021.

\bibitem{6998861}
P.~{Schniter} and S.~{Rangan},
\newblock ``Compressive phase retrieval via generalized approximate message
  passing,''
\newblock {\em IEEE Trans. Signal Process.}, vol. 63, no. 4, pp. 1043--1055,
  Feb. 2015.

\bibitem{8580585}
X.~{Meng} and J.~{Zhu},
\newblock ``Bilinear adaptive generalized vector approximate message passing,''
\newblock {\em IEEE Access}, vol. 7, pp. 4807--4815, 2019.

\bibitem{9293406}
Z.~{Yuan}, Q.~{Guo}, and M.~{Luo},
\newblock ``Approximate message passing with unitary transformation for robust
  bilinear recovery,''
\newblock {\em IEEE Trans. Signal Process.}, vol. 69, pp. 617--630, 2021.

\bibitem{guo2015approximate}
Q.~Guo and J.~Xi,
\newblock ``Approximate message passing with unitary transformation,''
\newblock {\em CoRR, vol. abs/1504.04799, 2015. [Online]. Available:
  http://arxiv.org/abs/1504.04799}.

\bibitem{9367220}
Z.~Shen, K.~Xu, and X.~Xia,
\newblock ``Beam-domain anti-jamming transmission for downlink massive {MIMO}
  systems: A stackelberg game perspective,''
\newblock {\em IEEE Trans. Information Forensics and Security}, vol. 16, pp.
  2727--2742, 2021.

\bibitem{6940305}
C.-K. Wen, S.~Jin, K.-K. Wong, J.-C. Chen, and P.~Ting,
\newblock ``Channel estimation for massive {MIMO} using gaussian-mixture
  bayesian learning,''
\newblock {\em IEEE Trans. Wirel. Commun.}, vol. 14, no. 3, pp. 1356--1368,
  Mar. 2015.

\bibitem{837052}
Da-Shan S., G.~J. Foschini, M.J. Gans, and J.M. Kahn,
\newblock ``Fading correlation and its effect on the capacity of multielement
  antenna systems,''
\newblock {\em IEEE Trans. Commun.}, vol. 48, no. 3, pp. 502--513, Mar. 2000.

\bibitem{7727995}
H.~Xie, F.~Gao, and S.~Jin,
\newblock ``An overview of low-rank channel estimation for massive {MIMO}
  systems,''
\newblock {\em IEEE Access}, vol. 4, pp. 7313--7321, 2016.

\bibitem{8477183}
S.~Beygi, A.~Elnakeeb, S.~Choudhary, and U.~Mitra,
\newblock ``Bilinear matrix factorization methods for time-varying narrowband
  channel estimation: Exploiting sparsity and rank,''
\newblock {\em IEEE Trans. Sig. Process.}, vol. 66, no. 22, pp. 6062--6075,
  Nov. 2018.

\bibitem{8322235}
R.~Zhang, H.~Zhao, and J.~Zhang,
\newblock ``Distributed compressed sensing aided sparse channel estimation in
  {FDD} massive {MIMO} system,''
\newblock {\em IEEE Access}, vol. 6, pp. 18383--18397, 2018.

\bibitem{6784090}
K.~{Huang} and N.~D. {Sidiropoulos},
\newblock ``Putting nonnegative matrix factorization to the test: a tutorial
  derivation of pertinent cramer—rao bounds and performance benchmarking,''
\newblock {\em IEEE Signal Process. Mag.}, vol. 31, no. 3, pp. 76--86, May
  2014.

\bibitem{6177982}
Y.~{Rong}, M.~R.~A. {Khandaker}, and Y.~{Xiang},
\newblock ``Channel estimation of dual-hop {MIMO} relay system via parallel
  factor analysis,''
\newblock {\em IEEE Trans. Wirel. Commun.}, vol. 11, no. 6, pp. 2224--2233,
  June 2012.

\bibitem{8879620}
Z.~{He} and X.~{Yuan},
\newblock ``Cascaded channel estimation for large intelligent metasurface
  assisted massive {MIMO},''
\newblock {\em IEEE Wirel. Commun. Lett.}, vol. 9, no. 2, pp. 210--214, Feb.
  2020.

\end{thebibliography}

\end{document}